\documentclass[aps,prb,twocolumn,showpacs]{revtex4}
\usepackage{amsfonts}
\usepackage{amscd}
\usepackage{bm}
\usepackage{amsmath}
\usepackage{amssymb}
\usepackage{mathptmx}
\usepackage{color}
\def\nicefrac#1#2{\frac{#1}{#2}}
\usepackage{ifpdf}
\usepackage[pdftex]{graphicx}
\DeclareGraphicsExtensions{.eps, .pdf, .jpg, .tif}
\usepackage{hyperref}
\usepackage{bookmath}
\usepackage{mathfrak}
\usepackage{mymatrices}
\usepackage{mycolors}
\def\root{\sqrt{2}}

\def\text#1{\mbox{\tiny #1}}
\def\sub#1{\mbox{\footnotesize #1}}
\def\orbital{{\tt SO(3)_{\text{L}}}}
\def\twoDorbit{{\tt SO(2)_{\text{L}_{\text{z}}}}}
\def\uniaxial{{\tt SO(2)_{\text{L}_{\text{z}}}} \times Z_2}
\def\spin{{\tt SO(3)_{\text{S}}}}
\def\gauge{{\tt U(1)_{\text{N}}}} 
\def\parity{{\tt P}} 
\def\time{{\tt T}}
\def\gaugeorbit{{\tt U(1)_{\text{L}_{\text{z}}- \text{N}}}} 
\def\mfp{\emph{mfp}}
\def\rf{\emph{rf}}
\def\oneeighth{\frac{\small 1}{\small 8}}
\begin{document}
\title{Chiral Phases of  Superfluid \He\ in an Anisotropic Medium}
\author{J.A. Sauls} 
\affiliation{Department of Physics and Astronomy, 
             Northwestern University, Evanston, Illinois 60208}
\date{September 18, 2012} %\date{\today}
\begin{abstract}
Recent advances in the fabrication and characterization of \emph{anisotropic} silica aerogels with
exceptional homogeneity provide new insight into the nature of unconventional pairing in disordered
anisotropic media. I report theoretical analysis and predictions for the equilibrium phases of
superfluid \He\ infused into a low-density, homogeneous uniaxial aerogel. Ginzburg-Landau (GL) theory for
a class of equal-spin-pairing (ESP) states in a medium with uniaxial anisotropy is developed and used
to analyze recent experiments on uniaxially strained aerogels. For \He\ in an axially ``stretched''
aerogel GL theory predicts a transition from normal liquid into a \emph{chiral} ABM phase at $T_{c_1}$ in
which the chirality axis, $\hvl$, is aligned \underline{along} the strain axis.
This orbitally aligned state, is protected from random fluctuations in the anisotropy
direction, has a positive NMR frequency shift, a sharp NMR resonance line and is identified with the
high-temperature ESP-1 phase of superfluid \He\ in axially stretched aerogel.
A second transition into a biaxial phase is predicted to onset at a slightly lower temperature,
$T_{c_2}<T_{c_1}$.
This phase is an ESP state, breaks time-reversal symmetry, and is defined by an orbital order
parameter that spontaneously breaks axial rotation symmetry. This transition is driven by the coupling of
an axially aligned 1D ``polar'' order parameter to the two time-reversed 2D axial ABM states.
The biaxial phase has a continuous degeneracy associated with the projection of its chiral axis
in the plane normal to the anisotropy axis. 
Theoretical predictions for the NMR frequency
shifts of the biaxial phase provide an identification of the ESP-2 as the biaxial state, partially
\emph{disordered} by random anisotropy (Larkin-Imry-Ma effect).
The ``width'' of the jump in the NMR frequency shift at $T_{c_2}$ provides an estimate of the orbital
domain size, $\xi_{\text{LIM}}\simeq 5\mu\mbox{m}$ at $18\,\mbox{bar}$.
I show that the random anisotropy results from mesoscopic structures 
in silica aerogels. This model for the random anisotropy field is coarse-grained on 
the atomic scale, and is formulated in terms of local anisotropy in the scattering of
quasiparticles in an aerogel with orientational correlations.
Long-range order of locally anisotropic scattering centers is related to the splitting of
the two ESP phases. 
\end{abstract}
\maketitle
%
%--------------------------------------------------------
The discovery of superfluidity of \He\ infused into high porosity silica
\emph{aerogel} opened a new chapter into the complex ordered phases of liquid
\He; and provided a novel method to study the effects of quenched disorder on the symmetry and
stability of unconventional pairing states in Fermi superfluids.\cite{por95,spr95}
The nature of the ordered phases of \He\ in high-porosity aerogel has recently been clarified 
% revealed
by experiments using silica aerogels with high homogeneity.\cite{pol08}

Nuclear magnetic resonance (NMR) spectroscopy has proven to be a powerful diagnostic of the symmetry of
the order parameter for the superfluid phases.\cite{leg75}
Pulsed NMR experiments on \He\ infused into a \emph{uniformly isotropic aerogel} with 98.2\%
porosity\cite{pol11} provide an unambiguous identification of the order parameter as a Balian-Werthamer
(BW) state,\cite{bal63} albeit with a significantly reduced order parameter amplitude,
$\Delta_{\text{B}}(T)$, and longitudinal resonance frequency, $\Omega_{\text{B}}(T)$, compared to that in
bulk \Heb.
Unlike the B-phase of pure \He\ which has a fully gapped excitation spectrum, analyses of heat
capacity,\cite{cho04a} thermal conductivity\cite{fis03} and magnetization measurements\cite{sau05} show
that the B phase of \Heaero\ is \emph{gapless}, with disorder-induced Andreev states dominating the
low-temperature thermal and transport properties.\cite{sha01}
The suppression of the A-phase and the absence of a poly-critical point (PCP) in isotropic aerogels
is consistent with theoretical predictions based on pair-breaking by scattering from a homogeneous
distribution of isotropic impurities, or a distribution of weakly anisotropic impurities with
orientational correlations on length scale, $\xi_{\sub{s}}$, less than the pair
correlation length, $\xi_0 = \hbar v_f/2\pi k_{\text{B}}T_c$.\cite{thu96,thu98,aoy06}

The fabrication of \emph{anisotropic} aerogels with exceptional homogeneity,\cite{pol08} have led to
recent superflow and NMR experiments that provide a clearer understanding of the nature of the superfluid
phases of \He\ in homogeneous silica aerogels.
Anisotropic stress acting on a homogeneously isotropic aerogel can dramatically alter the relative
stability of anisotropic pairing states by favoring one or more orbital components of the p-wave order
parameter.\cite{thu96,thu98,vic05,aoy06} 
Indeed the sensitivity of the order parameter to uniaxial strain is exhibited in
torsional oscillator experiments performed on \He\ confined in axially compressed aerogel.\cite{ben11}
These experiments show a large metastable region of the phase diagram upon cooling - \emph{assumed} to be
the chiral ABM state\cite{and60,bri73a} (A phase) - that extends well below the bulk PCP. Upon warming
from the low-temperature phase - presumed to be an anisotropic B phase - the transition into the
high-temperature A phase occurs at $T_{\text{AB}} = 0.075\,\mbox{mK}$ below the onset of superfluidity in
the aerogel ($T_{c_a}=2.275\,\mbox{mK}$ at $p=31.9\, \mbox{bar}$). This transition is absent for \He\ in
uncompressed isotropic aerogel, and is interpreted as the transition into an \emph{equilibrium} ABM state
stabilized by uniaxial strain.\cite{ben11}
However, identification of the high temperature superfluid phase as the ABM state is not
established.
Measurements of the superfluid density are insufficient to determine the symmetry of the order parameter.
Additional NMR experiments should clarify the symmetry of the ordered phases of \He\ in uniformly
anisotropic, compressed aerogels.

\begin{figure}[!]
\includegraphics[width=0.995\linewidth,keepaspectratio]{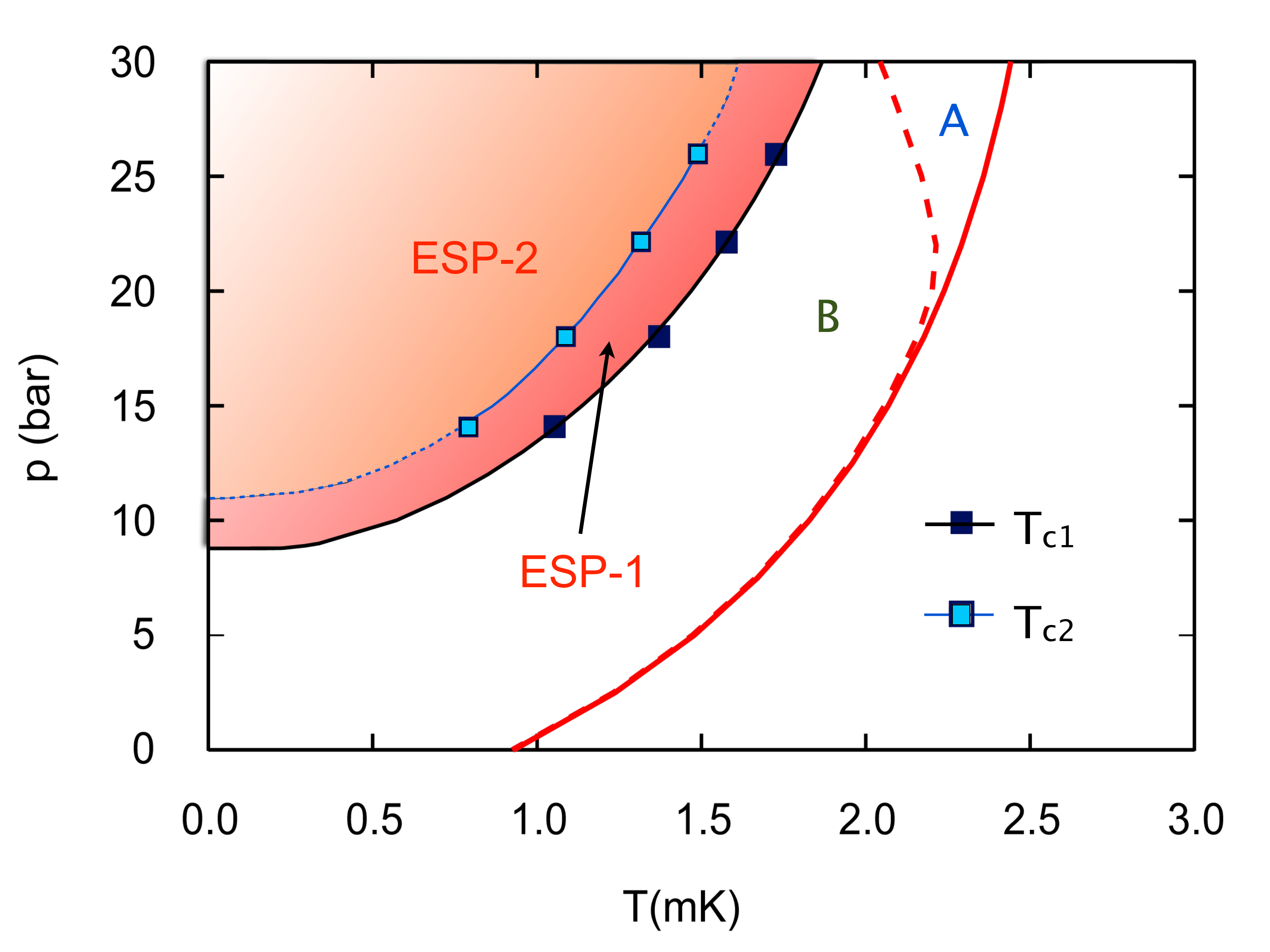}
\caption{\small
Two equal-spin-pairing phases labelled ESP-1 (pink region) and ESP-2 (orange region) identify the ordered
phases of \He\ in uni-axially stretched aerogel with 97.5\% porosity (adapted from Ref.
\onlinecite{pol12}). The transition at $T_{c_1}$ is substantially reduced below the bulk transition (red
line).
The second ESP state onsets at $T_{c_2}$.
An A-B transition at high pressure, characteristic of bulk \He\ (dotted red line), is not observed.
\label{fig-phase_diagram}
}
\end{figure}

NMR experiments on \He\ infused into uniformly anisotropic, ``axially stretched''
aerogel\footnote{``Stretched'' aerogels are not mechanically strained. Anisotropy equivalent to
uniform axial strain with $\varepsilon_{zz}>0$ is achieved by radial shrinkage during the drying process.
See Ref. \onlinecite{pol08}.} lead to a radically different phase diagram and interpretation of the
ordered phases than what is found for \He\ in isotropic aerogels.\cite{pol12}
Two distinct superfluid phases, both equal-spin pairing (ESP) states, are observed. The phase for
$T_{c_2}< T < T_{c_1}$ (ESP-1 in Fig. \ref{fig-phase_diagram}) was identified as an ABM state with the
chiral axis aligned \emph{perpendicular} to both the strain axis ($\hvz$) and the magnetic field, i.e.
the ``easy plane'' configuration with $\hvl \perp \hvz$. 
This identification is based on the observations of (i) a positive NMR frequency shift with
linewidth as narrow as the normal-state Larmor resonance, (ii) a tipping angle dependence of the NMR
shift in agreement with that predicted for the ABM state and (iii) a theoretical model
based on \emph{strain-alignment of random cylinders} to describe the local anisotropic structure of
aerogel.\cite{vol08}
This model is combined with the theory of Rainer and Vuorio\cite{rai77} for the orientation energy of the
chiral axis of \emph{pure} \Hea\ by a \emph{single} cylindrical impurity, which then predicts the chiral
axis to align \emph{with} the strain axis for compressed aerogels, i.e. ``easy axis'' with $\hvl ||
\hvz$ for $\veps_{\text{zz}}<0$, and \emph{perpendicular} to the strain axis for stretched aerogel, i.e.
``easy plane'' with $\hvl \perp \hvz$ for $\veps_{\text{zz}} > 0$.\cite{vol08,sur08,sur09}
This alignment model underlies the interpretation of the ordered phases 
for \He\ in uniaxially strained aerogels
reported by several groups.\cite{kun07,ben11,pol12}
However, the identification of uniaxial compression (stretching) with
``easy axis'' (``easy plane'') alignment of the chiral axis is model dependent - in this case upon
the alignment of rigid cylinders representing the local anisotropy of otherwise globally isotropic
aerogel and the assumption that the single impurity result of Ref. \onlinecite{rai77} extends to a
distribution cylindrical impurities with typical spacing, $\xi_a$, that is less than or the same
order as the pair correlation length, $\xi_0$.
Indeed the fractal structure of the aerogel\cite{meakin98,por99,haa01} on length scales shorter than
$\xi_a$ implies that the orientation of the chiral axis may not be inferred from the orientation energy
characteristic of a single cylindrical impurity in pure \He.
Furthermore, the response of a fractal network of silica strands and clusters to an external force
applied at the surface of an aerogel is a complex problem. The local stress distribution, changes
in bond angles, etc. may be very different from that based on rotation-alignment of rigid 
cylindrical impurities.\cite{ma00,ma02b}

%------------------------------------------------------------------------------------
%------------------------------------------------------------------------------------

In Sec. \ref{sec-GLfunctional} I step back from a microscopic description of how global anisotropy is
connected with local anisotropy and atomic forces, and consider the symmetry constraints for
the phases of superfluid \He\ embedded in a uniformly anisotropic medium on the scale of the pair
correlation length, $\xi(T)$, and develop a Ginzburg-Landau (GL) theory for ESP phases in such a medium.
These considerations lead to predictions for the order parameter and their NMR signatures for the phases
of superfluid \He\ infused into uniaxially stretched aerogel, as described in Secs.
\ref{sec-normal_2D}-\ref{sec-NMR_ESP-1}. 
In Secs. \ref{sec-normal_2D} and \ref{sec-NMR_ESP-1} I provide theoretical analysis for the
identification of the ESP-1 phase as an ABM state with the chiral axis aligned \underline{along} the
strain axis.
A second transition into a biaxial phase is predicted to onset at a slightly lower temperature,
$T_{c_2}<T_{c_1}$ (Sec. \ref{sec-biaxial_phase}).
This ESP phase breaks time-reversal symmetry, and is defined by a chiral orbital order
parameter that spontaneously breaks axial rotation symmetry. 
This transition is driven by the coupling of
an axially aligned 1D ``polar'' order parameter to the two time-reversed 2D axial ABM states.
The biaxial phase has a continuous degeneracy associated with the projection of its chiral axis
in the plane normal to the anisotropy axis. 
In Sec.
\ref{sec-NMR_ESP-2} I show that the NMR signatures of the ESP-2 phase are explained by a partially
disordered biaxial phase in which the chiral axis is disordered on a \emph{cone} - centered on the strain
axis - by the random anisotropy of the aerogel medium, the Larkin-Imry-Ma (LIM)
effect.\cite{lar70,imr75,vol08}
I also include a discussion of the possibility of a normal to 1D polar transition for \He\ in a strongly
anisotropic aerogel in Sec. \ref{sec-normal_1D}, as well as why this scenario does not explain the phase
diagram of Ref. \cite{pol12}
In Secs. \ref{sec-anisotropic_scattering_model}-\ref{sec-LRO_anisotropic_scattering} the GL theory is
combined with microscopic theory for pair breaking in a medium with both random and global anisotropy.
This allows for additional predictions connecting the normal-state transport properties of \He\ in
anisotropic aerogels with the symmetry of the superfluid phases.
In particular, in Sec. \ref{sec-LIM_anisotropic_scattering} I discuss a model for random anisotropy and 
the LIM effect within a scattering theory of partially ordered anisotropic impurities. This model is
the basis for the partially disordered biaxial phase that is identified with the ESP-2 phase.
\subsection{GL Theory for \He\ in Uniaxial Aerogel}\label{sec-GLfunctional}

The normal phase of pure liquid \He\ is separately invariant under spin- and orbital rotations, gauge
transformations ($\gauge$) as well as discrete symmetries of space ($\parity$) and time ($\time$)
inversion, \ie the group is ${\tt G} =
\spin\times\orbital\times\gauge\times\parity\times\time$.\footnote{The nuclear dipole-dipole interaction
breaks the relative spin- and orbital rotation symmetry, but is of order $E_{\text{D}}\simeq 10^{-7}\,\mbox{mK}$ per atom and can be treated perturbatively.\cite{leg75}}
The order parameter for pure \He\ belongs to vector representations of both spin and orbital rotation
groups, $\spin$ and $\orbital$.
The $2\times 2$ matrix representation for the spin-triplet, p-wave order
parameter,\cite{bal63,vollhardt90}
\ber
\Delta_{\alpha\beta}(\hvp) = \vec{\vd}(\hvp)\cdot\left(i\vec\vsigma\vsigma_y\right)_{\alpha\beta}
\,,
\eer
is parametrized by $\vec{\vd}(\hvp)$, which transforms as a vector under $\spin$, while the orbital
pairing states are a superposition of the $L = 1$ basis $\{\hp_i\,|\,i=x,y,z\}$ of 
$\orbital$, i.e. ${\vd}_{\alpha}(\hvp)=\sum_{i}A_{\alpha i}\,(\hvp_i)$.\cite{mer73}

The superfluid phases of \He\ in ``stretched'' aerogel are identified as ESP states.\cite{pol12} Here I
consider the class of ESP states of the form, $A_{\alpha i} = \vd_{\alpha}\,\cA_i$, where $\vec\vd$ is
a real unit vector in spin space orthogonal to the plane of the Cooper pair spins, and $\cA_i$ is a
complex vector under orbital rotations.

For \He\ embedded in a homogeneous, non-magnetic, anisotropic medium with inversion symmetry, the
orbital rotation symmetry is reduced, and the 3D basis, $\{\cA_i\,|\,i=x,y,z\}$, for the vector
representation of $\orbital$ is reduced to bases for 2D and 1D irreducible representations of
$\uniaxial$,
\be
\begin{pmatrix}\cA_x \cr \cA_y \cr \cA_z \end{pmatrix}
\xrightarrow[\text{strain}]{\text{uniaxial}}
{
\begin{pmatrix} a_x \cr a_y\end{pmatrix}\,\,\,
     \atop 
\begin{pmatrix} b \end{pmatrix}
	\,,
}
\ee
where $Z_2$ represents the identity and a $\pi$ rotation about an axis perpendicular to the uniaxial
strain axis (${\tt R}_{\pi}$), the latter denoted as $\hvz$. The 1D order parameter is the ``polar''
state, which is invariant under $\twoDorbit$ and changes sign for $b\xrightarrow[]{{\tt R}_{\pi}}
-b$.
The maximal symmetry group is then 
${\tt G}'=\spin\times\uniaxial\times\gauge\times\parity\times\time$.
For each irreducible representation there is a second-order invariant for the
leading order contribution to the Ginzburg-Landau functional.
For the class of ESP states there is one fourth order invariant for the 1D
representation, two fourth-order invariants for the 2D orbital representation and two mixed symmetry
invariants. Thus, the GL functional for zero magnetic field is
\ber\label{GLfunctional}
\Delta\Omega[\vec{a},b] &=& \alpha_{\perp}(T)\,\vert\vec{a}\vert^{2} 
                         +  \alpha_{\parallel}(T)\,\vert b\vert^{2}
\nonumber\\
                         &+& \beta_{1}\,\vert\vec{a}\vert^{4}
                          +  \beta_{2}\,\vert\vec{a}\cdot\vec{a}\vert^{2}
                          +  \beta_{3}\,\vert b\vert^{4}
                          +  \beta_{4}\,\vert\vec{a}\vert^{2}\,\vert b\vert^{2}
\nonumber\\
                         &+& \nicefrac{1}{2}\,\beta_{5}\,\left[\vec{a}\cdot\vec{a}\,(b^{*})^{2}
                                                   +  (\vec{a}\cdot\vec{a})^{*}\,b^{2}\right]
\,.
\eer

The coefficients of the second-order invariants determine the instability temperatures,
$T_{c_{\perp}}$ and $T_{c_{\parallel}}$, for the 2D and 1D order parameters, respectively. 
Thus, for temperatures $|T-T_{c_{\perp,\parallel}}| \ll T_{c}$ for unstrained aerogel,
$\alpha_{\perp,\parallel}(T)\simeq \alpha_{\perp,\parallel}^{'}
\left(T-T_{c_{\perp,\parallel}}\right)$, with $\alpha_{\perp,\parallel}^{'} > 0$.
The bare instability temperatures are equal in the isotropic limit, and thus for weak uniaxial anisotropy
we assume $T_{c_{\perp}} - T_{c_{\parallel}} = \lambda\,\varepsilon_{\text{zz}}\,T_{c}$, where the
uniaxial strain $\varepsilon_{\text{zz}} > 0$ ($\varepsilon_{\text{zz}} < 0$) for ``stretched''
(``compressed'') aerogel, and $\lambda$ is a material coefficient whose magnitude and \emph{sign} depend
on the microscopic mechanism by which anisotropy lifts the degeneracy between the 1D and 2D pairing
symmetry classes.

\subsection{Normal to 2D phases}\label{sec-normal_2D}

Consider the case in which $T_{c_{\perp}}>T_{c_{\parallel}}$. Thus, for $T\lesssim T_{c_{\perp}}$
there is necessarily a temperature region in which $\alpha_{\perp}<0$ and $\alpha_{\parallel}>0$. The
GL functional in Eq. \ref{GLfunctional} is then minimized with $b\equiv 0$ and reduces to
\be\label{GLperp}
\Delta\Omega[\vec{a}] = \alpha_{\perp}(T)\,\vert\vec{a}\vert^{2} 
                          +  \beta_{1}\,\vert\vec{a}\vert^{4}
                          +  \beta_{2}\,\vert\vec{a}\cdot\vec{a}\vert^{2}
\,.
\ee
The order parameter is a complex vector in the plane perpendicular to the strain axis ($\hvz$), and
is parametrized by
\be
\vec{a}=\Delta\,\left(\cos\varphi\,\hvx + e^{i\psi}\sin\varphi\,\hvy\right)
\,.
\ee
Minimizing the GL functional with respect to the amplitude $\Delta$ gives,
\be
\Delta\Omega = \onehalf\alpha_{\perp}(T)\,\Delta^2
\,,\quad\mbox{with}\quad
\Delta^2 = -\onehalf\frac{\alpha_{\perp}(T)}{\beta_1+\tilde{\beta}_2}
\,,
\ee
where $\tilde{\beta}_2(\varphi,\psi) = \beta_2(1 - \sin^2\psi\sin^2(2\varphi))$. Global stability
requires $\beta_1>0$ and $\beta_1+\beta_2>0$, however, there are two possibilities for the
equilibrium phase just below $T_{c_1}\equiv T_{c_{\perp}}$.
For $\beta_2 < 0$ the free energy is minimized for $\psi=0,\pi$ and \emph{any} $\varphi$, 
i.e. for an ``in-plane'' polar state,
\be\label{OP_ABM}
\vec{a}_{\text{P}} = \Delta_{\text{P}}\,\hvx
\,,\quad\mbox{with}\quad
\Delta_{\text{P}} = \sqrt{\onehalf\,\frac{\vert\alpha_{\perp}(T)\vert}{\beta_1+\beta_2}}
\,.
\ee
This phase preserves time-reversal symmetry, but spontaneously breaks the $\twoDorbit$ rotational
symmetry. The continuous degeneracy of the in-plane polar state under rotation of the polar axis in
the plane normal to the strain axis means that this phase will be subject to the Larkin-Imry-Ma (LIM)
effect; i.e. long-range orientational order of the in-plane polar order will be destroyed by random
fluctuations in the anisotropy direction.\cite{lar70,imr75,vol08}

For $\beta_2 >0$, which is the prediction of weak-coupling BCS
theory for weak anisotropic scattering,\cite{thu98} the free energy is minimized
for $\psi=\pm\pi/2$ and $\varphi=\pi/4$, i.e. by either of two degenerate \emph{chiral} ABM states,
\be
\vec{a}_{\text{ABM}}=\Delta_{\text{A}}\left(\hvx\, \pm i\,\hvy\right)/\root
\,,\quad\mbox{with}\quad
\Delta_{\text{A}} = \sqrt{\onehalf\,\frac{\vert\alpha_{\perp}(T)\vert}{\beta_1}}
\,.
\ee
The ABM state breaks time-reversal symmetry, but retains continuous axial symmetry, $\gaugeorbit$, by the
combining each element of $\twoDorbit$ with a gauge transformation. In contrast to the ABM state in pure
\He, the ABM phase stabilized by uniaxial anisotropy has its chiral axis, $\hvl=\pm\hvz$,
\underline{locked} parallel or anti-parallel to the strain axis. 
The absence of a continuous degeneracy associated with rotation of the chiral axis protects the ABM phase
against random fluctuations in the anisotropy direction.
%

%------------------------------------------------------------------------------------
\begin{figure}[h]
\includegraphics[width=0.995\linewidth,keepaspectratio]{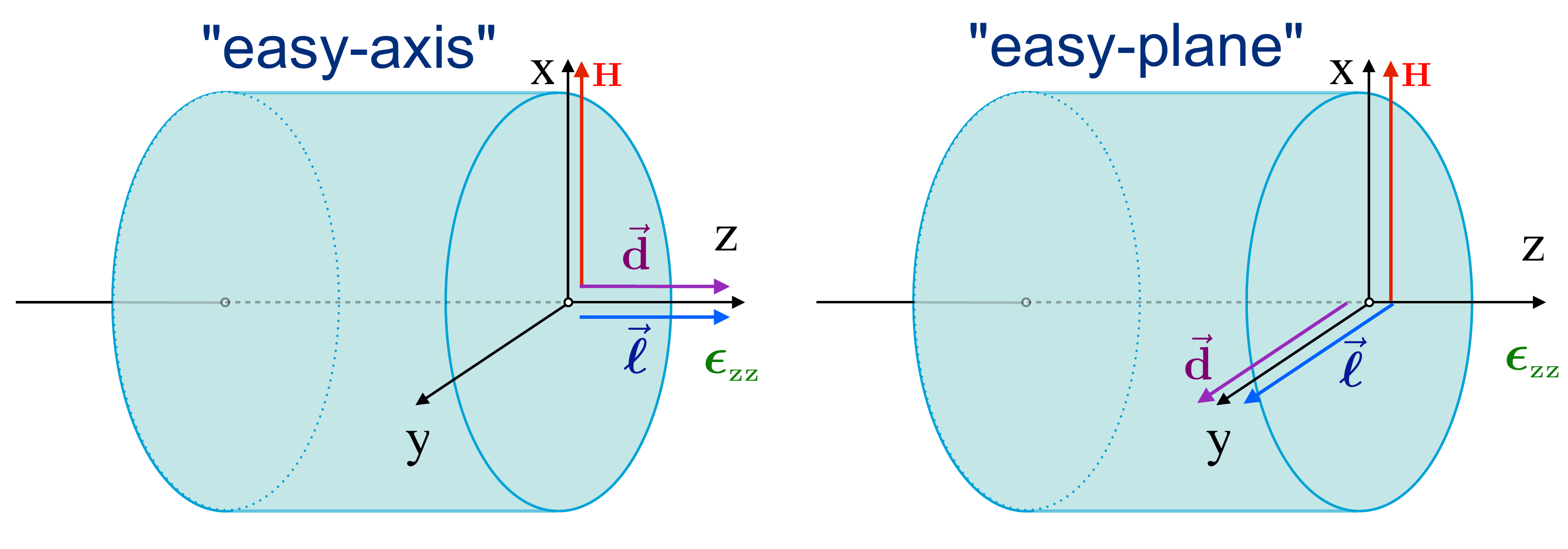}
\caption{\small
For ABM-phase confined in homogeneous uniaxial-strained aerogel the chiral axis is aligned
along the strain axis (``easy axis'' alignment) with $\hvl\,||\,\vz$.
Other models assign the chiral axis to the plane perpendicular to the strain axis (``easy plane''
alignment) with $\hvl\perp\vz$.
Dipole-locked configurations are shown for both cases shown with the magnetic 
field perpendicular to the strain axis, $\vH\,||\,\vx$.
\label{fig-aerogel_geometry}
}
\end{figure}
%------------------------------------------------------------------------------------

The geometry for the experiments reported in Ref. \onlinecite{pol12} is a cylinder of
\Heaero\ with uniaxial strain, $\veps_{\text{zz}}\simeq 0.14$, along the cylinder axis ($\hat\vz$), an
aspect ratio of $D/L \simeq 1.44$ and static magnetic field perpendicular to the strain axis, $\vH
\simeq 0.3 - 2.0\,\mbox{kG}\,\hat\vx$, as shown in Fig. \ref{fig-aerogel_geometry}.
The authors identified the ESP-1 phase for this ``stretched'' aerogel as a chiral ABM state with chiral
axis \emph{perpendicular} to the strain axis, the ``easy plane'' configuration shown
in Fig. \ref{fig-aerogel_geometry}.
This alignment is at odds with the GL theory presented here. 
A transition from the normal state to a chiral ABM state with $\hvl\perp\hvz$ is not allowed by symmetry.
In a uniform uniaxial medium a chiral ABM state is the equilibrium phase only if the chiral axis is
aligned with the strain axis, i.e. ``easy-axis'' alignment with $\hvl=\pm \hvz$.

The ABM state with $\hvl||\pm\hvz$ is also at odds with the model of \emph{strain-alignment of random
cylinders}\cite{vol08} combined with the alignment energy for a \underline{single} cylindrical
impurity,\cite{rai77} which is argued to favor ABM states with $\hvl\perp\hvz$ for an axially stretched
aerogel.
To emphasize the conflict of this model with GL theory, consider an ABM order parameter in a uniform
uniaxial medium - either stretched or compressed - with $\hvl\perp\hvz$. This requires the two
orbital amplitudes that define this ABM order parameter to belong to \emph{different} irreducible
representations of the maximal symmetry group, which implies that these two amplitudes onset with
\emph{different} instability temperatures. Thus, it is not possible for an ABM state with
$\hvl\perp\hvz$ to be the ESP-1 phase in a homogeneous uniaxial medium.
In the following Sec. \ref{sec-NMR_ESP-1} I identify the chiral ABM state with $\hvl||\pm\hvz$ with
the ESP-1 phase based on comparison between the theoretically expected NMR signatures and the NMR
measurements reported in Ref. \onlinecite{pol12}.

\subsection{NMR signatures for the ESP-1 Phase}\label{sec-NMR_ESP-1}

The Zeeman energy for an ESP state defined by $\vec{\vd}$ in a magnetic field, $\vec{\vH}$, 
is given by\cite{vollhardt90}
\be\label{Zeeman_Energy_A-phase}
\Delta\Omega_{\text{Z}} = g_{\text{z}}\,\Delta_{\text{A}}(T)^2\,\left(\vec\vd\cdot\vec{\vH}\right)^2
\,,
\ee
where $g_{\text{z}} > 0$.\cite{thu96} Thus, the ESP state accommodates the field by orienting the
$\vec{\vd}\perp\vec{\vH}$. The magnetization is then given by $\gamma\vec{\vS}=\chi\vec{\vH}$, where
$\gamma$ is the gyromagnetic ratio of \He\ and $\chi=\chi_{\text{N}}$ is the ABM-phase spin
susceptibility, which is unchanged from that of normal \He.\footnote{I neglect the particle-hole asymmetry correction of order $\delta\chi/\chi_{\text{N}}\simeq\Delta/E_f\simeq 10^{-3}$.}
When a pulsed transverse \rf\ field is applied along the strain axis, 
$\vec{\vH}_{1} = H_{1}\,\cos(\omega t)\hvz\perp\vec{\vH}$, it drives 
the magnetization away from the static field. Leggett's equation of
motion for the magnetization is
\ber\label{Leggett_S}
\partial_t\vec{\vS} 
&=& \gamma\left(\vec{\vS}\times\vec{\vH}\right) 
+ \gamma\left(\vec{\vS}\times\vec{\vH}_{1}\right) 
+ \vec{\vR}_{\text{D}} 
\,,
\\
\hspace*{-15mm}\mbox{where}\quad
\vec{\vR}_{\text{D}} 
&=& -\vec{\vd}
\times\left(\partial\Delta\Omega_{\text{D}}/\partial \vec{\vd}\right)
\,,
\label{dipole_torque}
\eer
is the torque from the nuclear dipole energy.\cite{leg75} The latter is defined by
\be\label{dipole_energy_ABM}
\Delta\Omega_{\text{D}} = g_{\text{D}}\,\Delta_{\text{A}}(T)^2\,\left[1-(\vec\vd\cdot\hvz)^2\right]
\,,
\ee
for an axially aligned ABM state.
The coupling constant $g_{\text{D}}>0$ is un-renormalized by impurity disorder,\cite{thu96}
and thus the dipole energy is minimized by aligning $\vec{\vd}||\hvl||\hvz$.
For time-dependent fields the $\vec{\vd}$ vector is driven by torque from the sum of the
external and internal fields,\cite{leg75}
\be\label{Leggett_d}
\partial_t\vec{\vd} 
= \gamma\,\vec{\vd}\times\left(\vec{\vH}-\frac{\gamma}{\chi}\vec{\vS}+\vec{\vH}_{1}\right) 
\,.
\ee
Note that for a purely static field the steady-state condition, $\gamma\vec{\vS}=\chi\vec{\vH}$, is
consistent with the minimum of the Zeeman energy, $\vec{\vd}\perp\vec{\vH}$. This condition is also
compatible with the orientation, $\vec{\vd}\perp\hvz$, which minimizes the dipole energy. This is
the ``easy axis'' geometry shown in Fig. \ref{fig-aerogel_geometry}.

For small \rf\ excitation $\vec{\vS}$ and $\vec{\vd}$ execute small excursions about
their equilibrium orientations, $\vec{\vS}_0 = S_0\hvx$ with $\gamma S_0 = \chi H$ 
and $\vec{\vd}_0=\hvz$, i.e. $\vec{\vS}=\vec{\vS}_0 + \delta\vec{\vS}$, $\vec{\vd}=\vec{\vd}_0 +
\delta\vec{\vd}$ with $\vec{\vd}_0\cdot\delta\vec{\vd}=0$. 
For the ``easy axis'' geometry the linearized equations
of motion reduce to
\ber
\partial_t\delta\vec{\vS} 
&=& \delta\vec{\vS}\times\vec{\omega}_{\,\text{L}} 
 + \vec{\vS}_{0}\times\vec{\omega}_{1} 
 - R_{\text{D}}\,\vec{\vd}_{0}\times\delta\vec{\vd}
\,,
\\
\partial_t\delta\vec{\vd} 
&=& -\frac{\gamma^2}{\chi}\vec{\vd}_{0}\times\delta\vec{\vS}
\,,
\eer
with $\vec{\omega}_{\,\text{L}}=\gamma\vec{\vH} = \omega_{\,\text{L}}\hvx$,
$\vec{\omega}_{1}=\gamma\vec{\vH}_{1}=\omega_{1}(t)\hvz$ and $R_{\text{D}}=
2g_{\text{D}}\,\Delta_{\text{A}}(T)^2$.
The latter determines
the longitudinal resonance frequency for the ABM-phase,
$\Omega_{\text{A}}^{2}=\gamma^2\,R_{\text{D}}/\chi$. 
For the ``easy axis'' geometry with the \rf\ field along the
strain axis the only components that are excited by the \rf\ field are $\delta{d}_x$, $\delta{S}_y$,
and $\delta{S}_z$. For a single Fourier component of frequency $\omega$ the coupled Leggett equations are
\be\label{NMR_dipole-locked_A}
\begin{pmatrix}         \omega       & -i\Omega_{\text{A}}   &      0                   \cr
		       +i\Omega_{\text{A}}   &     \omega            & -i\omega_{\,\text{L}}    \cr
				        0            & +i\omega_{\,\text{L}} &        \omega            \cr
\end{pmatrix}
\begin{pmatrix}\delta{D}_x \cr \delta{S}_y         \cr \delta{S}_z \end{pmatrix}
=	
\begin{pmatrix}    0       \cr -i\,S_0\,\omega_{1} \cr      0      \end{pmatrix}
\,,
\ee	
with $\delta{D}_x\equiv (R_{\text{D}}/\Omega_{\text{A}})\delta{d}_x$. The retarded 
linear response functions are then given by
\be\label{NMR_response_dipole-locked}
\begin{pmatrix}
	\delta{D}_x 	\cr 
	\delta{S}_y 	\cr 
	\delta{S}_z 
\end{pmatrix}
=
\frac{S_0\,\omega_{1}}{(\omega+i\eta)^2 - (\omega_{\,\text{L}}^2 + \Omega_{\text{A}}^2)}
\times
\begin{pmatrix}
	+\Omega_{\text{A}}		\cr
	-i\omega				\cr
	-\omega_{\,\text{L}}	 
\end{pmatrix}
\,,
\ee
with $\eta\rightarrow 0+$, which exhibit transverse NMR at $\omega = \sqrt{\omega_{\,\text{L}}^2 +
\Omega_{\text{A}}^2}$, and thus a maximum NMR frequency shift in the high-field limit,
$\omega_{\,\text{L}}\gg\Omega_{\text{A}}$,
\be\label{NMR_shift_ABM}
\Delta\omega=\omega-\omega_{\,\text{L}} \simeq \onehalf\Omega_{\text{A}}^2/\omega_{\,\text{L}}
\,.
\ee
For finite tipping angle, defined by
$\vec{\vS}(t=0+)=S_0(\cos\beta\hvx+\sin\beta\hvz)$, 
this result extends to the known result for the dipole-locked A-phase,\cite{bri75}
\be\label{A-phase_tipping_angle}
\Delta\omega = \onehalf\Omega_{\text{A}}^2/\omega_{\,\text{L}}
               \left( \nicefrac{1}{4} + \nicefrac{3}{4}\cos\beta \right)
\,.
\ee
These results for ``easy axis'' geometry agree quantitatively with experimental results for the ESP-1
phase in axially stretched aerogel reported in Ref. \onlinecite{pol12}, and thus the ESP-1 phase is
identified as the \emph{ABM state with chiral axis $\hvl$ aligned along the strain axis}.

A test of this identification can be made by re-orienting the ``stretched'' \Heaero\ sample
with the static field aligned along the strain axis, i.e. $\vec{\vH}=H\hvz$. In the high-field limit,
$\omega_{\,\text{L}}\gg\Omega_{\text{A}}$, this is a \emph{dipole-unlocked} configuration with
$\vec{\vd}_{0}\perp\vec{\vH}\leadsto\vec{\vd}_{0}\perp\hvl$, e.g. $\vec{\vd}_{0}=\hvy$ 
located at a maximum of the dipole potential.
Under \rf\ excitation with $\vec{\omega_{1}} || \hvx$ for small tipping angles the dipole 
torque is $\vec{\vR}_{\text{D}}=R_{\text{D}}\,\delta{d}_z\,\hvx$. The Leggett equations couple
$\delta{S}_x$, $\delta{S}_y$ and $\delta{D}_z=(R_{\text{D}}/\Omega_{\text{A}})\delta{d}_z$,
and lead to the response functions,
\be\label{NMR_response_dunipole-locked}
\begin{pmatrix}\delta{S}_x \cr \delta{S}_y         \cr \delta{D}_z \end{pmatrix}
\simeq
\frac{S_0\,\omega_{1}}{(\omega+i\eta)^2 - (\omega_{\,\text{L}}^2 - \Omega_{\text{A}}^2)}
\times
\begin{pmatrix}
-\omega_{\,\text{L}}\cr i\omega\cr -i\omega(\Omega_{\text{A}}/\omega_{\text{L}})
\end{pmatrix}
\,,
\ee
exhibiting a \emph{negative} frequency shift of the NMR resonance,
\be
\Delta\omega\simeq -\onehalf\Omega_{\text{A}}^2/\omega_{\,\text{L}}
\,.
\ee
By contrast, for the ``easy plane'' configuration of the chiral axis proposed in Ref. \onlinecite{pol12},
\underline{both} field orientations yield a positive frequency shift of
$\Delta\omega\simeq+\onehalf\Omega_{\text{A}}^2/\omega_{\,\text{L}}$.
Thus, these two orientations of the NMR field provide a stringent test of the identification of the ESP-1
order parameter, and in particular this GL theory of ESP pairing in a uniformly anisotropic medium.

\subsection{Interlude: Normal to 1D Transition}\label{sec-normal_1D}

Given that the onset of superfluidity in homogeneously anisotropic ``stretched'' aerogel reported in Ref.
\cite{pol12} is well described by a normal to 2D transition ($T_{c_{\perp}}>T_{c_{\parallel}}$) into the
axially aligned ABM state,
this theory predicts for the case, $T_{c_{\parallel}}>T_{c_{\perp}}$, a
normal to 1D transition.
In particular, the GL functional for $T\lesssim T_{c_{1}}\equiv T_{c_{\parallel}}$ is minimized by $\vec{a}\equiv 0$, and reduces to
\be\label{GLfunctional_polar}
\Delta\Omega[b] =  \alpha_{\parallel}(T)\,\vert b\vert^{2}
                          +  \beta_{3}\,\vert b\vert^{4}
\,.
\ee
The resulting equilibrium phase is an ESP state with an axially aligned polar 
order parameter,
\be
\vec{b}_{\text{P}}=b_{\text{P}}\,\hvz
\,,\quad\mbox{with}\quad
b = \sqrt{\onehalf\,\frac{\vert\alpha_{\parallel}(T)\vert}{\beta_{3}}}
\,.
\ee
Global stability requires $\beta_3>0$. 

The dipole potential for the ESP polar phase is
\be
\Delta\Omega_{\text{D}} = 2\,g_{\text{D}}\,b_{\text{P}}(T)^{2}\,(\vec{\vd}\cdot\hvz)^2
\,.
\ee
Thus, in contrast to the axially aligned ABM state, $\vec{\vd}\perp\hvz$ in equilibrium for the strain
aligned polar phase. For any other orientation of $\vec{\vd}$ the dipole torque is given by
\be
\vec{\vR}_{\text{D}}=-4\,g_{\text{D}}\,b_{\text{P}}(T)^{2}\,(\vec{\vd}\cdot\hvz)(\vec{\vd}\times\hvz)
\,.
\ee
The corresponding longitudinal resonance frequency for the polar phase is $\Omega_{\text{P}}^2 =
(\gamma^2/\chi)\,4 g_{\text{D}}\,b_{\text{P}}(T)^2$. Note that the ratio of the slopes of the
square of the longitudinal frequencies for the ABM and polar phases is given by
\be
\frac{\partial\Omega_{\text{P}}^2/\partial T\vert_{T_c}}
     {\partial\Omega_{\text{A}}^2/\partial T\vert_{T_c}}
=
2\frac{\alpha^{'}_{\parallel}}{\alpha^{'}_{\perp}}\frac{\beta_{1}}{\beta_{3}}
\,.
\ee
The Zeeman energy for the polar phase, 
\be\label{Zeeman_Energy_P-phase}
\Delta\Omega_{\text{Z}} = g_{\text{z}}\,b_{\text{P}}(T)^2\,\left(\vec\vd\cdot\vec{\vH}\right)^2
\,,
\ee
is minimized by $\vec{\vd}\perp\vec{\vH}$. Thus, for $\vH || \hvx$ both the dipole energy and Zeeman
energy are are minimized for $\vec{\vd}||\hvy$. 
Transverse \rf\ excitation with $\vec{\vH}_{1}||\hvz$ gives a transverse NMR resonance at
$\omega=\omega_{\,\text{L}}$, i.e. ``zero shift''. However, if we re-orient the static field along
the compression axis and the r.f. field transverse, e.g. $\vec\vH_1 || \hvx$, then we obtain positive NMR
shift of $\Delta\omega = \onehalf\Omega_{\text{P}}^2/\omega_{\,\text{L}}$ in the high field limit.
Note that the polar phase is also protected by the anisotropy energy from random fluctuations of the
anisotropy axis, and thus expected to exhibit a sharp NMR line for both orientations.

Recently \He\ has been infused into a new type of high-porosity aerogel formed from long strands of
aluminum oxide, so-called ``nematic aerogels'', exhibiting strong 
uniaxial anisotropy.\cite{ask12c} \He\ NMR
measurements indicate that the onset of superfluidity in this medium is to a 1D polar
phase.\cite{ask12} If this interpretation is correct then this is the first observation of a 1D polar
phase in superfluid \He. Added support for this interpretation is included in Sec.
\ref{sec-LRO_anisotropic_scattering}
\vspace*{-5mm}
\subsection{The 2$^{\text{nd}}$ ESP phase}\label{sec-biaxial_phase}

Returning to the normal to 2D case ($T_{c_{\perp}} >  T_{c_{\parallel}}$), for 
weak uniaxial anisotropy a second phase transition is predicted.
The ``bare'' instability temperature, $T_{c_{\parallel}}$, is
re-normalized by the 2D order parameter that develops below $T_{c_{\perp}}$.
Whether or not a second transition occurs depends on the magnitude of $T_{c_{\perp}}-T_{c_{\parallel}}$,
and the sign of the interaction terms coupling the 2D and 1D order parameters. In particular, a second
transition exists if the coefficient of the quadratic term for the 1D polar phase vanishes at $T_{c_2}$.
For the case in which the axially aligned ABM state (Eq. \ref{OP_ABM}) is the equilibrium state for
$T_{c_2}\le T < T_{c_1}$, the terms proportional to $\beta_5$ in Eq. \ref{GLfunctional} vanish at
$T_{c_2}$, in which case the 2$^{\text{nd}}$ order term in the GL functional for $T\rightarrow T_{c_2}$
is given by
\ber\label{GLfunctional-Tc2}
\Delta\Omega[\vec{a}_{\text{A}},b]  &=& \Delta\Omega_{\text{A}}(T) 
                          +  \tilde{\alpha}_{\parallel}(T)\,\vert b\vert^{2} + \cO(b^4)
\,,
\\
\hspace*{-15mm}\mbox{with}\quad
\tilde{\alpha}_{\parallel}(T) &=& \tilde{\alpha}_{\parallel}(T) + \beta_4\,\Delta_{\text{A}}(T)^2
\,,
\eer
where $\Delta\Omega_{\text{A}}(T)=\nicefrac{1}{2}\alpha_{\perp}(T)\,\Delta_{\text{A}}^2(T)$ is the free 
energy of the ABM phase.
Thus, a second-order instability to a mixed symmetry phase with $b\ne 0$ occurs for
$\tilde{\alpha}_{\parallel}(T_{c_2}) = 0$, which gives,
\be\label{Tc2}
T_{c_2} - T_{c_1} = 
\left(
\frac{\alpha^{'}_{\parallel}}{\alpha^{'}_{\parallel}-\alpha^{'}_{\perp}\,\beta_4/2\beta_1}
\right)
\left(T_{c_{\parallel}}-T_{c_{\perp}}\right)
\,.
\ee
For $\beta_4/\beta_1 > 0$ we have $T_{c_2} < T_{c_{\parallel}}$, but the transition is not
suppressed to zero temperature, at least within GL theory. 

Below $T_{c_2}$ the full GL functional in Eq. \ref{GLfunctional} must be minimized, including the
$\beta_5$ terms since the 2D order parameter need not remain a purely axial ABM phase. Indeed one expects
the 2D order parameter to be deformed from a pure ABM state due to the interactions terms. It is
convenient to parametrize the 2D order parameter below $T_{c_2}$ in terms of the chiral basis vectors,
$\hvx_{\pm} = \nicefrac{1}{\root}\left(\hvx \pm i \hvy\right)$,
\ber
\vec{a} = a_{+}\,\hvx_{+} + a_{-}\,\hvx_{-}
\,.
\eer
Note that $\hvx_{\pm}\cdot\hvx_{\pm}^* = 1$, $\hvx_{\pm}\cdot\hvx_{\pm} = 0$ and
$\hvx_{+}\cdot\hvx_{-}=1$. It is also useful to introduce amplitude and phase variables for each order
parameter,
\be
a_{\pm} = \Delta_{\pm}\,e^{i\alpha_{\pm}}\,,\quad b = \Delta_{z} \,e^{i\beta}
\,.
\ee
The GL functional can be conveniently expressed in terms of $u=\Delta_{+}^2$, $v=\Delta_{-}^2$, $w =
\Delta_{z}^2$ and the relative phase variable, $\Phi = \alpha_{+}+\alpha_{-} - 2\beta$,
\ber\label{GLfunctional2}
\Delta\Omega[u,v,w,\Phi] 
&=& \alpha_{\perp}(u+v) + \beta_1(u+v)^2 + \beta_2\,4\,u\,v
\nonumber\\
&+& \alpha_{\parallel}\,w + \beta_3\, w^2 
\nonumber\\
&+& \beta_4 (u+v)\, w
+ \beta_5 \sqrt{u}\sqrt{v}\,w\,\cos\Phi
\,.
\eer
The stationarity conditions with respect to $u$, $v$, $w$ and $\Phi$ are, 
\begin{widetext}
\ber
\label{GL1}
\pder{\Delta\Omega}{u} &=& 
\alpha_{\perp}
+2\beta_1 (u+v)+4\beta_2\,v 
+ \beta_4\,w 
+\onehalf\beta_5\sqrt{\frac{v}{u}}\,w\,\cos\Phi 
% &=& 0
=0
\,,
\hspace*{10mm}
\\
\pder{\Delta\Omega}{v} &=& 
\label{GL2}
\alpha_{\perp}
+2\beta_1 (u+v)+4\beta_2\,u 
+ \beta_4\,w 
+\onehalf\beta_5\sqrt{\frac{u}{v}}\,w\,\cos\Phi 
% &=& 0
=0
\,,
\\
\pder{\Delta\Omega}{w} &=&  
\label{GL3}
\alpha_{\parallel}
+2\beta_3 w + \beta_4\,(u+v) 
+ \beta_4\,w 
+\beta_5\sqrt{u}\sqrt{v}\,\cos\Phi 
% &=& 0
=0
\,,
\\
\pder{\Delta\Omega}{\Phi} &=& 
\label{GL4}
-\beta_5\sqrt{u}\sqrt{v}\,w\,\sin\Phi 
% &=& 0
=0
\,.
\eer
\end{widetext}
For $T < T_{c_2}$ we have $u\ne 0$ with both $v$ and $w$ growing continuously from zero at $T_{c_2}$. 
The difference of Eqs. \ref{GL1} and \ref{GL2} yields,
\ber
(v-u)\,\left\{4\beta_2 + \onehalf\beta_5\,\frac{w}{\sqrt{uv}}\,\cos\Phi\right\} = 0 
\,,
\eer
and since $v-u\ne 0$ and $w\ne 0$ for $T<T_{c_2}$ we obtain, 
\be\label{GL5}
\sqrt{uv} = -\frac{\small 1}{\small 8}\frac{\beta_5}{\beta_2}\,w\,\cos\Phi
\,.
\ee
This equation constrains the range of $\cos\Phi$: (i) $-1 \le \cos\Phi < 0$ for $\beta_5>0$, while
(ii) $0 < \cos\Phi \le 1$ for $\beta_5<0$. Eq. \ref{GL4} then fixes $\cos\Phi = \pm 1$ depending on the
sign of $\beta_5$.
The sum of Eqs. \ref{GL1} and \ref{GL2}, 
combined with Eq. \ref{GL5} gives 
\be
u + v = -\frac{\alpha_{\perp}}{2\beta_1} - \frac{\beta_4}{2\beta_1}\,w
\,,
\ee
which is used to obtain the free energy functional for the polar condensate density, $w$, 
for $T < T_{c_2}$,
\ber 
\Delta\Omega[w] 
&=& \Delta\Omega_{\text{A}}(T)  
                + \tilde{\alpha}_{\parallel}(T)\,w + \beta_{\text{w}}(\Phi)\,w^2
\,,
\\
\hspace*{-15mm}\mbox{where}\quad
\tilde{\alpha}_{\parallel}(T)
&=&\alpha_{\parallel}(T) - \onehalf\frac{\beta_4}{\beta_1}\alpha_{\perp}(T)
\\
\beta_{\text{w}}(\Phi)
&=& \beta_3 - \onefourth\frac{\beta_4^2}{\beta_1}-\oneeighth\frac{\beta_5^2}{\beta_2}\cos^2\Phi
\,.
\eer
A second transition develops when $\tilde{\alpha}_{\parallel}(T_{c_2})= 0$, which is the instability
temperature given in Eq. \ref{Tc2}. Below $T_{c_2}$ the polar phase density is given by
\be
w(T) = -\frac{\tilde{\alpha}_{\parallel}}{2\beta_{\text{w}}(\Phi)}
\,,
\ee
and a reduced thermodynamic potential given by,
\be
\Delta\Omega(T) = -\onefourth\frac{\alpha_{\perp}^2}{\beta_1} 
                  - \onefourth\frac{\tilde{\alpha}^2_{\parallel}}{\beta_{\text{w}}(\Phi)}
\,,\quad T< T_{c_2}
\,.
\ee
The ordered state for $T < T_{c_2}$ with the lowest free energy is given by the smallest allowed positive
value of $\tilde{\beta}_{\text{w}}(\Phi)$, which fixes the internal phase to be $\cos\Phi = +1$ for
$\beta_5 < 0$ or $\cos\Phi = -1$ for $\beta_5 > 0$. 

The orbital order parameter, above and below $T_{c_2}$, can be expressed in the form,
\be
\vec{a}_{\text{A}} = \Delta_{\text{A}}(T)\,\hvx_{+}\,,\quad T_{c_2}<T<T_{c_1}
\,,
\ee
where I have assigned the phase $\alpha_{+}=0$ for the axially aligned ABM phase.
The order parameter for the low-temperature mixed-symmetry phase then takes the form,
\be\label{biaxial_OP}
\hspace*{-2.5mm}\vec{\cA} 
= \Delta_{+}(T)\,\hvx_{+}
+ e^{i\alpha_{-}}\Delta_{-}(T)\,\hvx_{-}
+ e^{i\beta}\Delta_{z}(T)\,\hvz
\,,\,T<T_{c_2}\,,
\ee
where the order parameter amplitudes are given by,
\ber
\Delta_{\text{A}}(T)
&=& 
\bar{\Delta}_{\text{A}}\,\sqrt{1-T/T_{c_1}}
\,,\quad T \le T_{c_1}
\,,
\label{ABM_gap}
\\
\bar{\Delta}_{\text{A}} 
&=&
\sqrt{
\frac{T_{c_{1}}}{2\beta_1}\,\der{\alpha_{\perp}}{T}\Big\vert_{T_{c_{1}}}
}
\,.
\eer
for the axially aligned ABM phase above $T_{c_2}$ (ESP-1).
For the low temperature phase
the polar amplitude is given by
\ber
\Delta_{z} 
&=&
\bar{\Delta}_{z}\,\sqrt{1-T/T_{c_2}}
\,,\quad T \le T_{c_2}
\,,
\\
\bar{\Delta}_{z} 
&=&
\sqrt{
\frac{T_{c_2}}{2\beta_1}\,\der{\tilde{\alpha}_{\parallel}}{T}\Big\vert_{T_{c_2}}
\,
\frac{1}{\bar{\beta}_{\text{w}}}
}
\,,
\\
\quad\mbox{with}\quad
\bar\beta_{\text{w}}
&=&
\bar\beta_3-\nicefrac{1}{4}\,\bar\beta_4^2-\nicefrac{1}{8}\,\bar\beta_5^2/\bar\beta_2
\,.
\eer
Note the dimensionless $\beta$-parameters are normalized by $\beta_1$; $\bar\beta_{i}=\beta_{i}/\beta_1$.
For $T<T_{c_{2}}$ we can express the two chiral amplitudes, $\Delta_{\pm}(T)$, in terms of
$\Delta_{z}(T)$ and Eq. \ref{ABM_gap} for 
$\Delta_{\text{A}}(T)$ - extended to $T< T_{c_{2}}$ - as follows: 
\ber\label{Delta_pm}
\Delta_{\pm}&=&\nicefrac{1}{2}(\Delta_{s}\pm\Delta_{d})
\,,
\\
\,\nonumber\\
\label{Delta_s}
\Delta_{s} 
&=&\sqrt{
\Delta_{\text{A}}^2(T)-\onehalf(\bar{\beta}_4
-\onehalf\bar{\beta}_5/\bar{\beta}_2)\,\Delta^2_{z}(T)
}
\,,
\\
\label{Delta_d}
\Delta_{d} 
&=&\sqrt{
\Delta_{\text{A}}^2(T)-\onehalf(\bar{\beta}_4
+\onehalf\bar{\beta}_5/\bar{\beta}_2)\,\Delta^2_{z}(T)
}
\,.\hspace*{10mm}
\eer

\subsubsection*{Bi-axial and Chiral Order}

The low-temperature ESP phase has a continuous degeneracy associated with the \emph{relative} phases of
the polar ($\beta$) and ABM amplitudes ($\alpha_{\pm}$) defining the mixed symmetry order parameter in
Eq. \ref{biaxial_OP}. For $\bar\beta_{5}>0$ the constraint $\alpha_{+}+\alpha_{-} - 2\beta = \pi$, allows
us to parametrize the internal degeneracy by a single phase angle on the interval, 
$0 \le \varphi \le 2\pi$, and express the mixed-symmetry order
parameter as
\be\label{biaxial_OP_gauge-orbit}
\vec{\cA}
= \Delta_{+}(T)\,e^{-i\varphi}\,\hvx_{+}
- \Delta_{-}(T)\,e^{+i\varphi}\,\hvx_{-}
+ \Delta_{z}(T)\,\hvz
\,,
\ee
up to an overall phase. Note that the chiral basis vectors transform under gauge transformations as
\be\label{gauge-rotation}
e^{\mp i\varphi}\,\hvx_{\pm} = \hvx_{\pm}^{'} = \left(\hvx^{'} \pm i \hvy^{'}\right)/\root
\,,
\ee
where $\hvx^{'}=\cos\varphi\,\hvx + \sin\varphi\,\hvy$ and $\hvy^{'}=-\sin\varphi\,\hvx +
\cos\varphi\,\hvy$ are rotations of the in-plane orbital axes, ($\hvx,\hvy$), about the strain axis,
$\hvz$. This implies that the continuous degeneracy associated with the internal phase, $\varphi$,
corresponds to \emph{spontaneous breaking of the axial symmetry} of the ABM phase at $T_{c_2}$.

Further insight is obtained by considering the product, $\cA_i\,\cA_j^*$. This tensor determines
observables such as the the momentum dependent ``gap function'', $\Delta(\hvp)$, the superfluid density
tensor, $(\rho_{s})_{ij}$, the nuclear dipole-dipole energy, $\Delta\Omega_{\text{D}}$, the intrinsic
angular momentum density, $\vec{\cL}$,
etc.
The tensor $\cA_i\,\cA_j^* = \Delta_{ij} + \cL_{ij}$
separates into a real symmetric tensor,
$\Delta_{ij}$, and an imaginary, anti-symmetric tensor, $\cL_{ij}$. 
The real order parameter tensor is given by,
\ber\label{OP_tensor_real}
\Delta_{ij} 
  &=& \nicefrac{1}{2}\left(\Delta_{+}^2+\Delta_{-}^2\right)\,
                \left(\hvx_{i}\hvx_{j}+\hvy_{i}\hvy_{j}\right) + \Delta_{z}^2\,\hvz_{i}\hvz_{j}
\nonumber\\
  &-&\Delta_{+}\Delta_{-}\,\left(\hvx_{i}\hvx_{j}-\hvy_{i}\hvy_{j}\right)
\nonumber\\
  &+&\nicefrac{1}{\root}\,\Delta_{z}(\Delta_{+}-\Delta_{-})\,\left(\hvx_{i}\hvz_{j}+\hvz_{i}\hvx_{j}\right)
\,.
\eer
The first line of terms in Eq. \ref{OP_tensor_real}, which preserve axial symmetry, includes the polar
distortion, $\sim\hvz_{i}\hvz_{j}$, that varies as $\Delta_{z}^2\sim(1-T/T_{c_2})$ below
$T_{c_2}$. However the second line,
$\sim\left(\hvx_{i}\hvx_{j}-\hvy_{i}\hvy_{j}\right)$, exhibits the spontaneously
broken axial symmetry of the ESP-2 phase. Thus, the ESP-2 phase possesses \emph{bi-axial anisotropy} 
with a magnitude scaling as $\Delta_{+}\Delta_{-}\sim(1-T/T_{c_2})$ below $T_{c_2}$.
The polar and in-plane distortions conspire to generate the bi-axial anisotropy represented
by $\sim\left(\hvx_{i}\hvz_{j}+\hvz_{i}\hvx_{j}\right)$, which 
scales as $\Delta_{z}(\Delta_{+}-\Delta_{-})\sim(1-T/T_{c_2})$ below
$T_{c_2}$. 
The biaxial anisotropy of the ESP-2 phase can be expressed in terms of semi-major, $\Delta_{s}\hvy$, and
semi-minor, $\Delta_{d}\hvx$, axes defining the in-plane gap distortion by combining the terms using Eqs.
\ref{Delta_pm}-\ref{Delta_d}. The continuous degeneracy of the bi-axial phase corresponds to the
orientation of the semi-major and semi-minor axes in the plane perpendicular to the strain axis $\hvz$.
Since $\Delta_{ij}$ is real and symmetric, it can be expressed in diagonal form in
terms of tensor products of an orthonormal triad, $\{\hvm,\hvn,\hvl\}$, that can be expressed 
as a rotation of the laboratory axes, $\{\hvx,\hvy,\hvz\}$, as follows,
\ber
\hvm &=& +\cos\vartheta(\cos\varphi\,\hvx + \sin\varphi\,\hvy) + \sin\vartheta\,\hvz
\,,\\
\hvn &=& -\sin\varphi\,\hvx + \cos\varphi\,\hvy
\,,\\
\hvl &=& -\sin\vartheta(\cos\varphi\,\hvx + \sin\varphi\,\hvy) + \cos\vartheta\,\hvz
\,.
\eer
where $0\le \varphi\le 2\pi$ is the in-plane gauge-rotation angle defined in Eq.
\ref{biaxial_OP_gauge-orbit} that parametrizes the degeneracy of the bi-axial phase, while the polar
rotation, $\vartheta$, is fixed by energetics,
\ber\label{cone_angle}
\cos\vartheta = \Delta_{d}/\sqrt{\Delta_{d}^2+2\Delta_{z}^2}
\,,\,\,
\sin\vartheta= \root\Delta_{z}/\sqrt{\Delta_{d}^2+2\Delta_{z}^2}
\,.
\eer
The resulting biaxial tensor order parameter reduces to 
\ber\label{OP_tensor_real3}
\Delta_{ij} &=& \nicefrac{1}{2}\,\left(\Delta_{d}^2+2\Delta_{z}^2\right)\,\hvm_{i}\hvm_{j}
             +  \nicefrac{1}{2}\,\Delta_{s}^2\,\hvn_{i}\hvn_{j} 
\,.
\eer
The zero eigenvalue associated with the eigenvector $\hvl$ means that $\pm\hvl$ are \emph{nodal
directions} of the momentum-space pair amplitude, $|\Delta(\hvp)| =
(\hp_{i}\,\Delta_{ij}\,\hp_{j})^{\nicefrac{1}{2}}$.
Thus, the nodal points associated with the ABM state (ESP-1 phase) are not destroyed by the second-order
transition to the bi-axial ESP-2 phase, but \emph{rotate} from the points $\pm\hvz$ (corresponding to
momenta along the strain axis) to the points $\pm\hvl$.
This rotation of the point nodes off the $\hvz$ axis leads directly to the continuous degeneracy of the
biaxial phase characterized by the \emph{orientation} of the nodal points in the plane perpendicular to
the strain axis as shown in Fig. \ref{fig-biaxial_NMR_geometry}. Thus, $\hvl$ defines the spontaneously
broken axis appearing below $T_{c_2}$, whose degeneracy is parametrized by the gauge-rotation angle
$\varphi$.

Furthermore, the nodal directions reflect the \emph{chiral} nature of the bi-axial phase. 
This is revealed by the anti-symmetric order parameter tensor, $\cL_{ij}$, which can be 
expressed as,
\ber\label{OP_tensor_imaginary}
\cL_{ij} 
= -\nicefrac{i}{2}\,\Delta_{s}\Delta_{d}\,\varepsilon_{ijk}\,\hvz_{k}
  +\nicefrac{i}{2}\root\Delta_{z}\Delta_{s}\,\varepsilon_{ijk}\,\hvx_{k}
\,.
\eer
For $T>T_{c_2}$ $\cL_{ij}\rightarrow\nicefrac{-i}{2}\Delta_{\text{A}}\,\varepsilon_{ijk}\hvz_{k}$, which
is directly related to the \emph{intrinsic angular momentum density},
$\vec{\cL}_{\text{A}}=\kappa_{a}(4m/\hbar)\Delta_{\text{A}}^2\,\hvz$, for a condensate of Cooper pairs
each with orbital angular momentum $+\hbar$ along $\hvz$.\cite{leg75,cro75,vol81,cho89}
Below $T_{c_2}$ $\cL_{ij}$ can also be expressed in terms of a single chiral axis, 
\ber\label{OP_tensor_imaginary2}
\cL_{ij} 
= -\nicefrac{i}{2}\,\Delta_{s}\,\sqrt{\Delta_{d}^2+2\Delta_{z}^2}\,\varepsilon_{ijk}\,\hvl_{k}
\,.
\eer
generating the nodal directions along $\pm\hvl$, and an intrinsic angular momentum density
in the bi-axial phase given by
\be
\vec{\cL} = \kappa_{a}(4m/\hbar)\Delta_{s}\sqrt{\Delta_{d}^2+2\Delta_{z}^2}\;\hvl
\;.
\ee
The GL material coefficient $\kappa_a$ was calculated by Choi and Muzikar.\cite{cho89} For pure \He\ the
resulting intrinsic angular momentum density is exceedingly small, $\cL\sim
n(\Delta(T)/E_f)^2\,\hbar$, where $n$ is the \He\ density. However, impurity disorder leads to
larger orbital currents, reflected in $\kappa_{a}\sim (n/E_f)\,\xi_0^2\,(\xi_0/\bar\ell)$ where
$\xi_0=\hbar v_f/2\pi T_c$ is the Cooper pair size and $\bar\ell$ is the transport mean-free path
resulting from scattering by impurities. This leads to an intrinsic angular momentum density of order 
$\cL\sim n\hbar(\xi_0/\bar\ell)(\Delta(T)/2\pi T_c)^2$.\cite{cho89}
Experimental observation of the intrinsic angular momentum density would provide a direct signature of
\emph{chiral order} predicted for both ESP phases discussed here for \He\ in uniaxially stretched
aerogel.

\subsection{NMR signatures of the ESP-2 phase}\label{sec-NMR_ESP-2}

The recent report of the discovery of two \emph{chiral} superfluid phases of \He\ in uniaxially stretched
areogel\cite{pol12} is based on their NMR signatures. The identification of the ESP-1 as the axially
aligned ABM phase was discussed in Sec. \ref{sec-NMR_ESP-1}.
The authors of Ref. \onlinecite{pol12} tentatively identified the ESP-2 phase s a \emph{textural
transition} of an ``easy-plane'' ABM state.
However the theory presented here demonstrates that this identification is \underline{not} allowed by
symmetry for superfluid \He\ infused into a \emph{uniformly anisotropic} aerogel.

Here I calculate the NMR signatures of the biaxial phase predicted for $T\le T_{c_2}$, and compare with
the observed NMR spectra for the low temperature (ESP-2) phase.
This leads to the identification of the ESP-2 phase as a ``biaxial LIM phase'' resulting from
orientational disorder induced by the random potential, i.e. the Larkin-Imry-Ma (``LIM'') 
effect,\cite{lar70,imr75} discussed in the context of \Heaero\ by Volovik.\cite{vol96,vol08} 
First consider the spin dynamics for a \emph{homogeneous} biaxial phase.

The Zeeman energy takes the same form as Eq. \ref{Zeeman_Energy_A-phase} for the axially aligned ABM,
\ber\label{Zeeman_Energy_biaxial_phase}
\Delta\Omega_{\text{Z}} 
&=& g_{\text{z}}\,\Delta_{\text{B}}(T)^2\,\left(\vec\vd\cdot\vec{\vH}\right)^2
\,,
\eer
but with the order parameter amplitude % replaced by
\ber
\label{gap_biaxial_phase}
\Delta_{\text{B}}^2 
&=& \Delta_{\text{A}}^2 
 + \left(1-\nicefrac{1}{2}\bar\beta_4\right)\Delta_{z}^2
\,.
\eer

%----------- Biaxial OP, LIM State and NMR geometry ----------
\begin{figure}[h]
\includegraphics[width=0.995\linewidth,keepaspectratio]{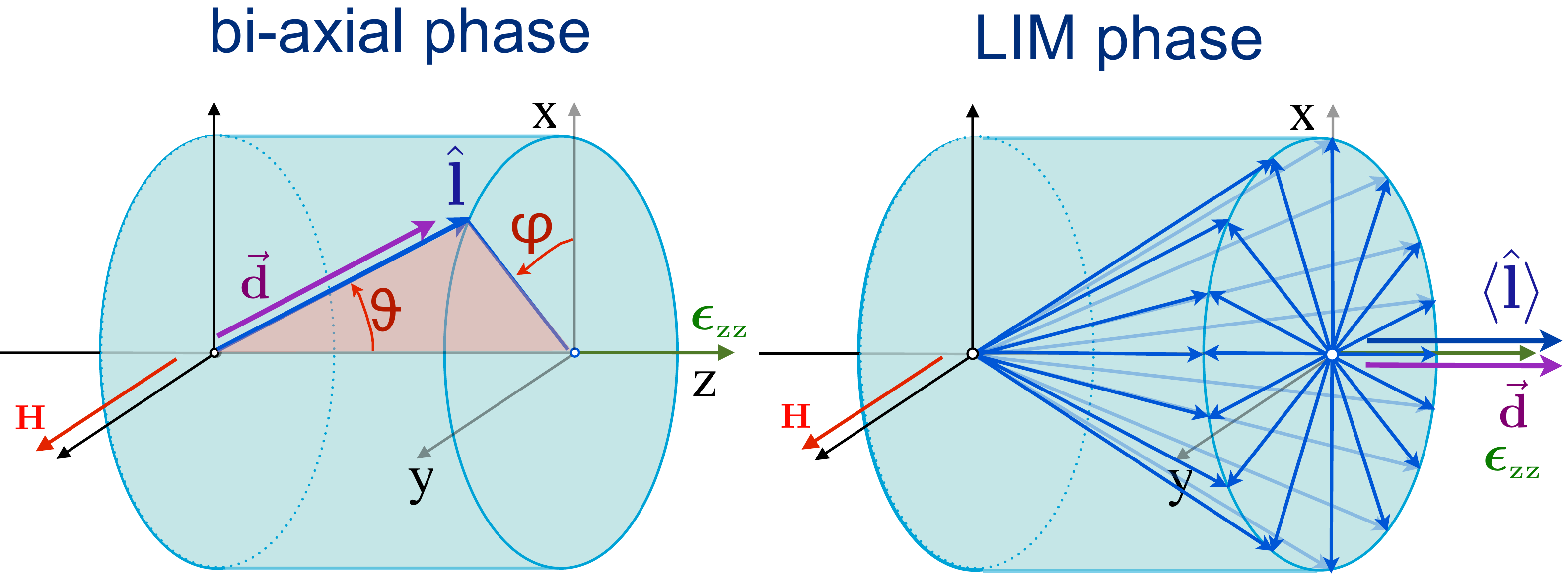}
\caption{\small
Left: The biaxial phase in uniaxially stretched aerogel is represented by the chiral
axis, $\hvl$, which can lie on a cone with angle $\vartheta$ relative to the strain axis, $\hvz$. 
The dipole energy is minimized by $\vec{\vd}\parallel\pm\hvl$.
Right: For the disordered biaxial LIM phase the chiral axis is distributed on the ``degeneracy cone''.
The average dipole energy is minimized for $\vec{\vd}\parallel\pm \langle\hvl\rangle$, where
$\langle\hvl\rangle=\langle\vl\rangle\,\hvz$ is the LIM-averaged chiral order parameter.
\label{fig-biaxial_NMR_geometry}
}
\end{figure}
%------------------------------------------------------------------------------------

The nuclear dipolar potential, 
$\Delta\Omega_{\text{D}}=2g_{\text{D}}\,\vd_{i}\,\Delta_{ij}\,\vd_{j}$, is
simply expressed in the basis of biaxial eigenvectors, $\{\hvm,\hvn,\hvl\}$,
\ber\label{Dipole_Energy_biaxial_phase}
\Delta\Omega_{\text{D}} 
&=&
g_{\text{D}}\,\left(\Delta_{d}^2+2\Delta_{z}^2\right)\,(\hvm\cdot\vec{\vd})^2
 + 
g_{\text{D}}\,\Delta_{s}^2\,(\hvn\cdot\vec{\vd})^2
\,.
\eer
The dipole energy is minimized by orienting $\vec\vd$ parallel to the chiral axis $\hvl$, the latter
of which is degenerate in orientation on a cone centered about the $\hvz$ axis defined by angle
$\vartheta$ as shown in Fig. \ref{fig-biaxial_NMR_geometry}. 
For the homogeneous biaxial phase the Zeeman
energy resolves the continuous degeneracy by orienting the dipole-locked biaxial state with
$\vec\vd\parallel\hvl\perp\vec\vH$, i.e. $\varphi=0,\pi$.
For any other orientation of $\vec{\vd}$ the dipole torque is given by
\be
\hspace*{-3mm}
\vec{\vR}_{\text{D}} =
- 2g_{\text{D}} 
\left[
\left(\Delta_{d}^2+2\Delta_{z}^2\right)(\vec{\vd}\cdot\hvm)\vec{\vd}\times\hvm
+ 
\Delta_{s}^2 (\vec{\vd}\cdot\hvn)\vec{\vd}\times\hvn
\right]
\,.
\ee
The linearized Leggett equations are of the same form as Eqs. \ref{Leggett_S} and \ref{Leggett_d},
but with $\vec{\vd}_0=\hvl$ and $\vec{\vR}_{\text{D}}$ given by
\ber
\vec\vR_{\text{D}} &=& \overleftrightarrow\vR_{\text{D}}\cdot\left(\vec\vd_{0}\times\delta\vec\vd\right)
\,,
\\
\hspace*{-5mm}\mbox{where}\quad
\overleftrightarrow\vR_{\text{D}} &=& -R_{s}\left(\hvm\,\hvm + \hvl\,\hvl\right) - R_{d}\,\hvn\,\hvn
\,,
\eer
$R_{s}=2g_{\text{D}}\Delta_{s}^2$ and $R_{d}=2g_{\text{D}}(\Delta_{d}^2+2\Delta_{z}^2)$.

For the geometry in Fig. \ref{fig-biaxial_NMR_geometry}, $\gamma\vec\vS_0 = \chi\vH$,
$\vec{\omega}_{\,\text{L}}=\gamma\vec{\vH} = \omega_{\,\text{L}}\hvy$, and
$\vec{\omega}_{1}=\gamma\vec{\vH}_{1}$,
and for the dipole-locked biaxial state with $\vec\vd_0=\hvl\perp\vec\vH$, the chiral axis is
confined in the $x-z$ plane, and the biaxial triad can be expressed in the NMR coordinates:
$\hvm\equiv\hvx'=\cos\vartheta\hvx-\sin\vartheta\hvz$, 
$\hvn=\hvy$ and 
$\hvl\equiv\hvz'=\cos\vartheta\hvz+\sin\vartheta\hvx$.
In this basis the linearized equations separate into a pair of equations,
for the longitudinal response,
\be\label{biaxial_NMR_response_longitudinal}
\begin{pmatrix} \delta D_{x'} \cr \delta S_{y} \end{pmatrix}
	=
	-\frac{(-i R_{d}/\Omega_{d})(\hvy\cdot\vec\omega_{1})}{(\omega+i\eta)^2 - \Omega_{d}^2}
\begin{pmatrix} \omega \cr -i\Omega_{d} \end{pmatrix}
\ee
with $\delta D_{x'}\equiv R_{d}\delta\vd_{x'}/\Omega_{d}$ and 
\be
\Omega_{d}^2 \equiv \frac{\gamma^2}{\chi}\,2g_{\text{D}}(\Delta_{d}^2+2\Delta_{z}^2)
\,.
\ee
The pole at $\omega = \Omega_{d}$ is the longitudinal NMR resonance frequency, which is 
excited only if $\hvy\cdot\vec\omega_{1}\ne 0$.
The transverse spin response functions are given by 
\ber\label{biaxial_NMR_response_transverse}
\hspace*{-15mm}
\begin{pmatrix}
	\delta{S}_{x'} 	\cr 
	\delta{D}_{y} 	\cr 
	\delta{S}_{z'} 
\end{pmatrix}
&=&
\frac{\chi/\gamma^2}{(\omega+i\eta)^2 - (\omega_{\,\text{L}}^2 + \Omega_{s}^2)}
% \times
\nonumber\\
&\times&
% &&
\begin{pmatrix}
	+i\omega\omega_{\text{L}}(\hvz'\cdot\vec\omega_1) 
	- (\omega_{\,\text{L}}^2 + \Omega_{s}^2)(\hvx'\cdot\vec\omega_1)	\cr
    +\Omega_{s}\left[\omega_{\text{L}}(\hvz'\cdot\vec\omega_1) 
	+i\omega (\hvx'\cdot\vec\omega_1) \right]							\cr 
    -\omega_{\,\text{L}}\left[\omega_{\text{L}}(\hvz'\cdot\vec\omega_1) 
	+i\omega (\hvx'\cdot\vec\omega_1) \right] 	
\end{pmatrix}
\,,
\eer
with $\Omega_{s}^2\equiv(\gamma^2/\chi)\,2g_{\text{D}}\Delta_{s}^2$ given by
\be
\Omega_{s}^2
= 
(\gamma^2/\chi)\,2g_{\text{D}}
\left(\Delta_{\text{A}}^2  
         -\nicefrac{1}{2}(\bar\beta_4-\nicefrac{1}{2}\bar\beta_5/\bar\beta_2)\Delta_{z}^2
\right)
\,.
\ee
The transverse resonance is at $\omega = \sqrt{\omega_{\,\text{L}}^2 + \Omega_{s}^2}$, with a maximum
positive NMR frequency shift in the high-field limit, $\omega_{\,\text{L}}\gg\Omega_{s}$, given by
\be
\hspace*{-3mm}
\Delta\omega \simeq \nicefrac{1}{2}\,\Omega_{s}^2/\omega_{\text{L}}
= \frac{\gamma^2}{\chi\omega_{\text{L}}}\,g_{\text{D}}
   \left(\Delta_{\text{A}}^2  
         -\nicefrac{1}{2}(\bar\beta_4-\nicefrac{1}{2}\bar\beta_5/\bar\beta_2)\Delta_{z}^2
   \right)
\,.
\ee
The predicted transverse shift is continuous at $T=T_{c_2}$, i.e.
$\Delta\omega\vert_{T_{c_2}}=\nicefrac{1}{2}\Omega_{\text{A}}^2/2\omega_{\text{L}}\vert_{T_{c_2}}$, but
with a discontinuity in slope, $\partial\Delta\omega/\partial T\vert_{T_{c_2}}$, governed by the polar
distortion, $\Delta_{z}$, and GL coefficients, $\bar\beta_{2,4,5}$. For
$\bar\beta_4-\nicefrac{1}{2}\bar\beta_5/\bar\beta_2 >0$ we expect the polar distortion to lead to a
reduction in the slope of the NMR shift below $T_{c_2}$.

%----------- NMR Shifts for ESP-1 and ESP-2 phases -------------------
\begin{figure}[h]
\includegraphics[width=0.995\linewidth,keepaspectratio]{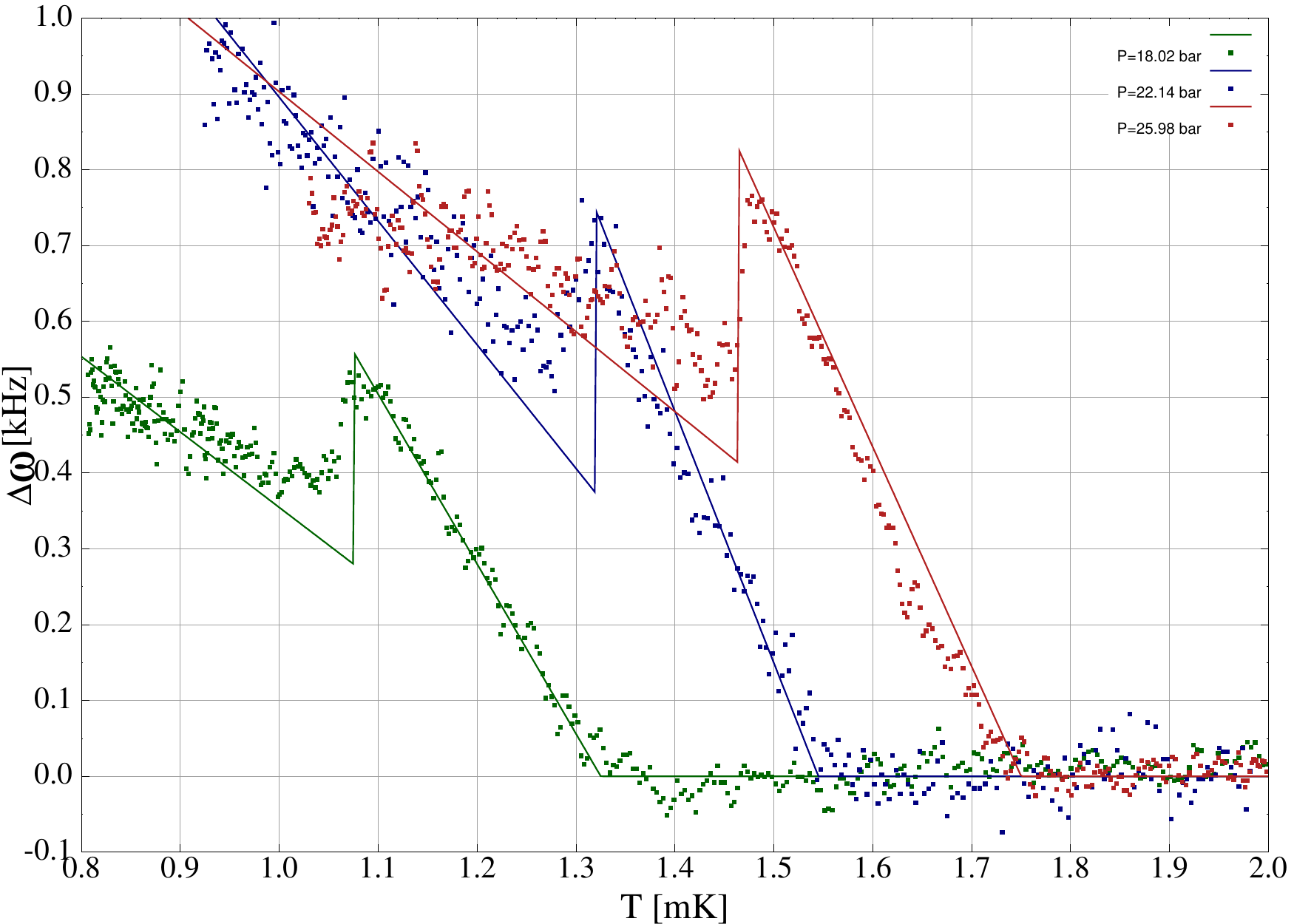}
\caption{\small
NMR frequency shifts for both ESP-1 and ESP-2 phases reproduced from Ref. \onlinecite{pol12}
for pressures: 
$18\,\,\mbox{bar}$ [\darkgreen{\tiny $\blacksquare$}],  
$22\,\,\mbox{bar}$ [\blue{\tiny $\blacksquare$}] and 
$26\,\,\mbox{bar}$ [\darkred{\tiny $\blacksquare$}].  
Theoretical curves for the same pressures are based on the predicted NMR shifts for the axially aligned
ABM state (ESP-1) and the biaxial LIM phase (ESP-2) with a predicted negative jump of $\nicefrac{1}{2}$
at $T_{c_2}$. The slopes are fits that are consistent with the theoretically predicted temperature
dependences for both ESP-1 and ESP-2 phases.
\label{fig-NMR_shifts}
}
\end{figure}
%------------------------------------------------------------------------------------

However, as shown in Fig. \ref{fig-NMR_shifts} the data for the NMR shifts of \He\ in uniaxially
stretched aerogel and reported by Pollanen et al.\cite{pol12} shows a more dramatic temperature
dependence for the frequency shift below the second transition: a negative ``jump'' identified with
$T_{c_2}$, followed by an increasing shift below $T_{c_2}$, but with a reduced slope compared to the ESP-1
phase above $T_{c_2}$.
In addition, the NMR linewidth (shown in Fig. 3 of Ref. \onlinecite{pol12}), which is sharp and virtually
unchanged in the axially aligned ABM phase (ESP-1), increases rapidly below $T_{c_2}$. Both the jump in
$\Delta\omega$ and the increased linewidth suggest that the ESP-2 phase exhibits some form of
\emph{orbital disorder} responsible for inhomogeneous broadening of the NMR spectrum and a reduction in
the first moment (the shift). 
Below I consider orbital disorder, induced by random anisotropy, and its effects on the NMR spectrum 
for the biaxial phase.

First consider the axially aligned ABM phase above $T_{c_2}$ for which the chiral axis is 
aligned along the strain axis, $\hvl = \hvz$.
There is \emph{no continuous orientational degeneracy} in the direction of the chiral axis.
Thus, fluctuations in the local anisotropy axis of the aerogel contribute to the suppression of
$T_c$ and the magnitude of the order parameter, but long-range orientational order is preserved because
there is a finite energy cost to long-wavelength transverse fluctuations of the chiral axis.

\vspace*{-6mm}
\subsubsection*{Effect of Random Anisotropy on the Bi-axial Phase}
\vspace*{-2mm}

By contrast the biaxial phase has a \emph{continuous rotational degeneracy} corresponding to orientation
of the chiral axis, $\hvl$, \emph{on a cone} with angle $\vartheta$ fixed by the polar component of the
order parameter (Eq. \ref{cone_angle}) as shown Fig. \ref{fig-biaxial_NMR_geometry}.
Fluctuations in the local anisotropy of the aerogel medium couple to the components of
$\hvl$ transverse to the strain axis, and destroy long-range orientational order of
the chiral axis\cite{lar70,imr75,vol08} - more precisely the \emph{transverse} components of the chiral
axis.
In particular, the random-field averages of the bi-axial triad are
\be
\langle\hvm\rangle = \sin\vartheta\,\hvz\,,\quad
\langle\hvn\rangle = 0\,,\quad
\langle\hvl\rangle = \cos\vartheta\,\hvz
\,,
\ee
and the correlation function for the transverse components of the chiral axis,
$\delta\hvl=\hvl-\langle\hvl\rangle$, 
\be
\langle\,\delta\hvl_{i}(\vr)\delta\hvl_{j}(\vr')\rangle 
\approx
\nicefrac{1}{2}
\sin^2\vartheta\,(\delta_{ij}-\hvz_{i}\hvz_{j})
\;e^{-|\vr-\vr'|^2/2\xi_{\text{LIM}}^2}
\,.
\ee
exhibit short-range order up to a length scale, $\xi_{\text{LIM}}$, that depends on the microscopic 
model for the random anisotropy field and its coupling to the orbital order parameter (see Sec.
\ref{sec-anisotropic_scattering_model}).
Nevertheless, different models for the random anisotropy field lead to orbital domain sizes
that are typically smaller than the dipole coherence length.

In the limit $\xi_{\text{D}}\gg\xi_{\text{LIM}}$ the spin-orbit coupling of $\vec\vd$ and $\hvl$, and
thus the NMR frequency shift, average to zero for a \emph{globally isotropic}
aerogel.\cite{thu96,vol96}
However, for the biaxial state in a \emph{globally anisotropic} aerogel the random anisotropy field
averaging leads to a spin-orbit coupling of $\vec\vd$ and $\langle\hvl\rangle \parallel\hvz$,
\ber
\langle\Delta\Omega_{\text{D}}\rangle 
=
&-& 2g_{\text{D}}\,
\left(\nicefrac{1}{2}\Delta_{\text{A}}^2-(1+\nicefrac{1}{4}\bar\beta_4)\Delta_{z}^2\right)\,
\left(\vec{\vd}\cdot\hvz\right)^2
\,,
\eer
that is the same form as
that for the axially aligned ABM phase (ESP-1), resulting in a transverse shift,
\be\label{NMR_shift_LIM_biaxial}
\Delta\omega 
= \frac{\gamma^2}{\chi\omega_{\text{L}}}\,g_{\text{D}}
   \left(\nicefrac{1}{2}\Delta_{\text{A}}^2-(1+\nicefrac{1}{4}\bar\beta_4)\Delta_{z}^2\right)
\,,
\ee
which is \emph{half} the transverse shift of the ESP-1 phase for $T\rightarrow T_{c_2}-$. Thus, the
orbitally disordered biaxial phase exhibits a ``negative jump'' of the NMR shift to half that of the ESP-1
NMR shift and a reduction in the slope of the shift for $T<T_{c_2}$. This result is in good agreement
with the observed temperature dependence of the NMR shifts reported by Pollanen et al.\cite{pol12} as
shown in Fig. \ref{fig-NMR_shifts}.
The theoretical slopes are fit to the experimental results, and 
are in agreement with theoretical expectations based on Eqs. \ref{NMR_shift_ABM}
and \ref{NMR_shift_LIM_biaxial}.

The predicted jump of $\nicefrac{1}{2}$ is the maximum reduction in the shift resulting from the
averaging the triad of orbital vectors over a cone with fixed polar angle, $\vartheta$.
The magnitude of the jump is reduced if the transverse components $\delta\hvl$ are not uniformly
distributed on the cone shown in Fig. \ref{fig-biaxial_NMR_geometry}.
Furthermore, since $\sin\vartheta\sim\Delta_z(T)\rightarrow 0$ as $T\rightarrow T_{c_2}$ we have
$\hvl\rightarrow \hvz$. Thus, averaging on the collapsing cone becomes irrelevant sufficiently close to
$T_{c_2}$, and we must recover the full EPS-1 NMR shift continuously over a narrow temperature ``width''
of order $\delta T_{c_2}\simeq T_{c_2}(\xi_{\text{LIM}}/\xi_{D})^2\ll T_{c_2}$ as the cone angle closes
towards the axially aligned ABM phase (see Sec. \ref{sec-LIM_anisotropic_scattering}).
Such a cross-over very close to $T_{c_2}$ is visible in Fig. \ref{fig-NMR_shifts} with observed
``widths'' of order $\delta T_{c_2}\approx 0.1\,\mbox{mK}$, providing an estimate for the orbital domain
size of $\xi_{\text{LIM}}\approx \nicefrac{1}{3}\xi_{\text{D}}\approx 5\,\mu\mbox{m}$ at
$18\,\mbox{bar}$.
Orbital domains of size $\xi_{\text{LIM}}\lesssim \xi_{\text{D}}$ also provide a
plausible explanation for the onset of increased NMR linewidth observed just below $T_{c_{2}}$.
But are such large domains of the orbital order parameter plausible? 

Volovik addressed the issue of orbital disorder in superfluid \He\ induced by random
anisotropy\cite{vol96,vol08} and derived a formula for the domain size based on arguments similar to
those of Larkin,\cite{lar70} Imry and Ma.\cite{imr75}
Volovik's result for the LIM correlation length,\cite{vol08}
\be\label{LIM-Volovik}
\xi_{\text{LIM}}^{\text{Volovik}} = \xi_{a}(\xi_0/d)^2
\,,
\ee
is based on (i) anisotropy derived from randomly oriented cylinders of mean spacing, $\xi_{a}\approx
20\,\mbox{nm}$, and diameter, $d\approx 3\,\mbox{nm}$, representing the aerogel, and (ii) the
orientational energy for a single cylindrical impurity in bulk \Hea\ calculated by Rainer and
Vuorio, $E_{a} \approx T_c\,k_f^2\,\xi_{a}\,d$.\cite{rai77}
This gives an orbital correlation length $\xi_{\text{LIM}}^{\text{Volovik}} \lesssim
1\,\mu\mbox{m}$ - weakly pressure dependent and an order of magnitude or more smaller than the
dipole coherence length, $\xi_{\text{D}}$ (black curves in Fig. \ref{fig-LIM_size}).
The ratio $\xi_{\text{LIM}}^{\text{Volovik}}/\xi_{\text{D}}$ decreases dramatically at lower pressures. 
Implicit in this calculation is the assumption that the single impurity result of Ref. \onlinecite{rai77}
extends to a distribution cylindrical impurities with typical spacing, $\xi_{a}$, that is less than
or the same order as the pair correlation length, $\xi_{0}$.
Indeed the fractal structure of the aerogel on length scales \emph{shorter} than $\xi_{a}$ may 
be responsible for weaker local anisotropy and thus a larger orbital domain size than 
the estimate from Eq. \ref{LIM-Volovik}.\cite{sur08}

\subsection{Anisotropic Scattering Model}\label{sec-anisotropic_scattering_model}

%------------------------------------------------------------------------------------
\begin{figure}[t]
\includegraphics[width=0.99\linewidth,keepaspectratio]{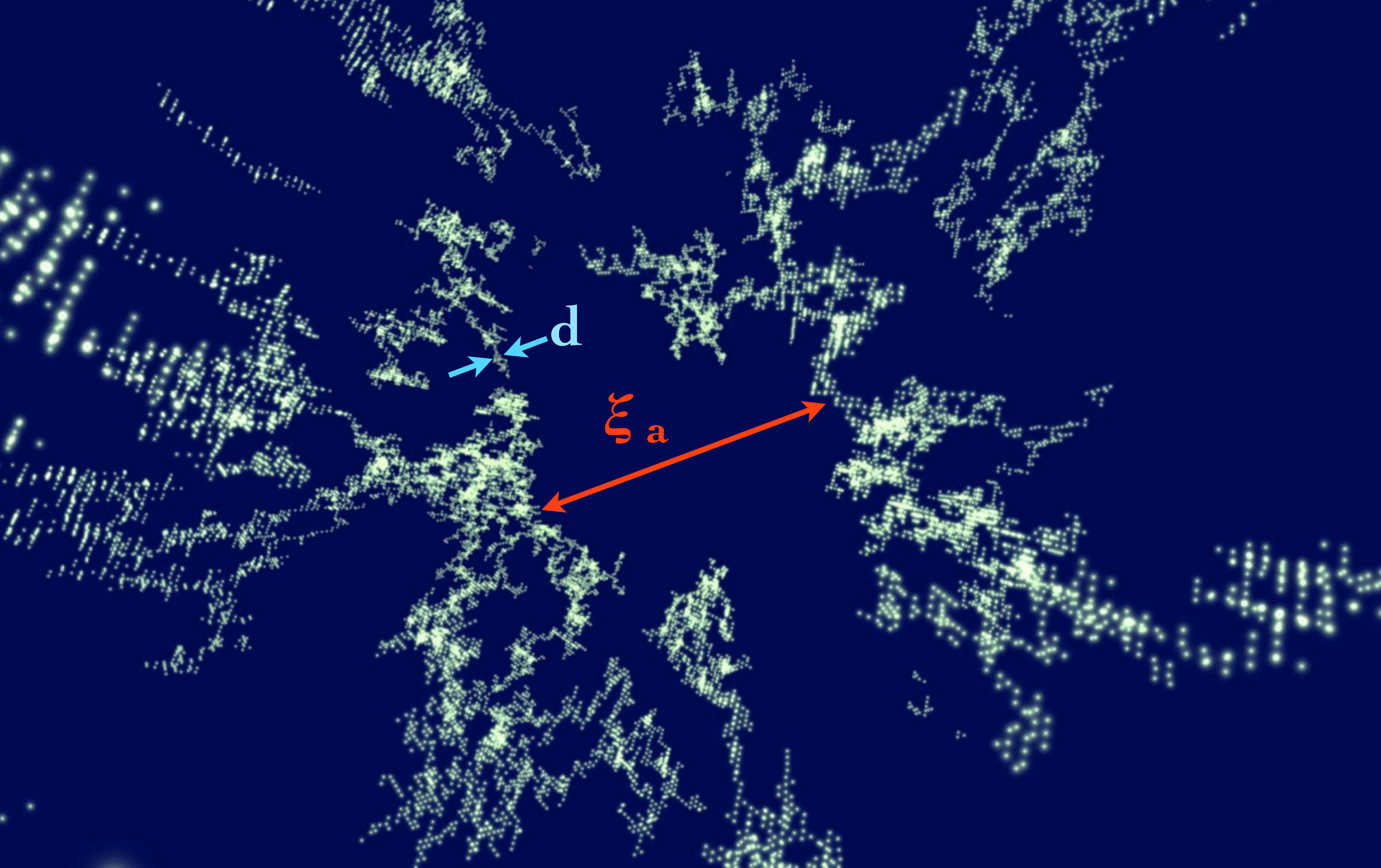}
\caption{\small
Local anisotropy of silica clusters and strands based on a DLCA simulation for the
growth of a 98\% porous aerogel.
Statistical self-similarity is observable over three decades of length scales.
The aerogel correlation length is of order $\xi_a \simeq 30\,\mbox{nm}$, while the strand
size is of order $d\simeq 2\,\mbox{nm}$.
\label{fig-dlca_aerogel}
}
\end{figure}
%------------------------------------------------------------------------------------

Silica aerogels grow by gelation of silica clusters. The resulting structure factors
measured by SAXS on high porosity silica aerogels are in good agreement with numerical simulations based
on diffusion-limited-cluster aggregation (DLCA).\cite{meakin98,por99}
A DLCA simulation of 98\% aerogel is presented in Fig. \ref{fig-dlca_aerogel} showing structures that are
locally anisotropic, as well as statistically self-similar over several decades in length scales. The
aerogel correlation length, $\xi_a$, is characteristic of the largest ``void'' dimension, while the
microscopic scale is indicated by the ``strand'' width, $d\ll\xi_a$, in Fig. \ref{fig-dlca_aerogel}.
Note that local anisotropy results from clusters and strands with multiple length scales.

In the following I formulate a model of the random field in aerogels, including random anisotropy, in
terms of the distribution of ballistic paths for quasiparticles propagating through the open regions of
aerogel. Elastic scattering by the fractal structure limits ballistic propagation. The local
cross-section, or scattering rate, is then the measure of random anisotropy. Such a description was
discussed by Thuneberg et al.\cite{thu96,thu98} as a course-grained model of random anisotropy compared
to the atomic scale $d$ of a silica strand. Below I expand on this model.

The scattering of quasiparticles by the aerogel is formulated in terms of the amplitude, $u(\vp,\vp')$,
for quasiparticle transitions from $\vp\rightarrow\vp'$ for a random distribution of scattering centers
(``strands'' or ``clusters'') with average density $n_s$. At low temperatures, $T\ll E_f$, the elastic 
scattering rate is
\be
\frac{1}{\tau_{\vp,\vp'}} \equiv w(\vp,\vp')=\pi\,n_s\,N_f\,|u(\vp,\vp')|^2
\,,
\ee
where $N_f$ is the quasiparticle density of states at the Fermi energy.
If the scattering medium is locally isotropic the scattering rate 
may be expanded in Legendre functions, e.g. $w(\vp,\vp')=w_0 + w_1\,\hvp\cdot\hvp' + \ldots$,
where I include s- and p-wave scattering. 
However, the scattering centers are anisotropic. On mesoscopic length scales, $d
\ll \delta r < \xi_a$ the anisotropy is locally well defined, and the scattering rate will depend on
the directions of the incident and scattered quasiparticle momenta \emph{relative} to a set of anisotropy
axes defining a local region of scattering centers.
This is illustrated by considering a medium of randomly distributed, but identical cylindrical
``strands''. The scattering medium is then locally uniaxial and the scattering rate is determined by the
local orientation of the anisotropy axis of the ``strand'', defined by $\hvs$.\footnote{The
``strand'' model describes scattering by \emph{strands} or \emph{disks}.}
For this ``strand model'' the scattering rate in the s-p approximation,
\ber\label{strand_model}
\frac{1}{\tau_{\vp,\vp'}} = w_0 + \hvp_{i}\,w_{ij}\,\hvp'_j
\,,
\eer
is parametrized by an isotropic scattering rate, $w_0$, and a uniaxial tensor,
\be\label{uniaxial_scattering_matrix}
w_{ij}= w_{\perp}\,(\delta_{ij} - \hvs_{i}\hvs_{j})
      + w_{\parallel}\,\hvs_{i}\hvs_{j}
\,,
\ee
with p-wave scattering rates, $w_{||}$ and $w_{\perp}$, for scattering preferentially along the symmetry
axis and perpendicular to the symmetry axis, respectively.
The random anisotropy field for the strand model is then encoded in the distribution of of the local
anisotropy axis $\hvs(\vr)$.\footnote{A more general model for random anisotropy would
also include the rate parameters, $w_0$, $w_{\perp},w_{\parallel}$, as random variables since the
structure of aerogel is far from a model of identical impurities. There are, however, constraints on the
rate parameters. For example, the scattering rate is non-negative for all relative orientations of the
incoming and scattered momenta. This constraint implies $w_0>0$ and $w_0 \pm w_{\perp,\parallel} \ge 0$.
Furthermore, the local scattering centers are typically \emph{not} uniaxial, but biaxial, $w_{ij} =
w_{s}\,\hvs_{i}\hvs_{j} + w_{s}\,\hvt_{i}\hvt_{j} + w_{u}\,\hvu_{i}\hvu_{j}$, and described by three
principal anisotropy axes, $\hvs,\hvt,\hvu$ and corresponding rate parameters.}
For a \emph{globally isotropic medium} the anisotropic scatters remain oriented over a finite   
correlation length, $\xi_{s}$, defined by the decay of the orientational correlations,
\be\label{strand_correlations_short-range}
\langle\hvs(\vr)\cdot\hvs(0)\rangle \sim e^{-r^2/2\xi_s^2}
\,,
\ee
where the configurational average can be defined in terms of a joint probability 
distribution for the orientation of all impurities. 
A reasonable estimate for this correlation length based on the DLCA simulations is the
aerogel correlation length, $\xi_{a}\approx 30-50\,\mbox{nm}$.

The relative scale of the orientational correlations to that of the pair correlation length, $\xi_0$, is
an important parameter. For $\xi_s\gg\xi_0$, scattering by the aerogel structure leads to
anisotropic pair breaking effects on the orbital states of p-wave Cooper pairs, and a splitting of the
superfluid transition for Cooper pairs with orbital motion parallel and perpendicular to the anisotropy
direction, $\hat{\vs}$, \ie $T_{c\perp}\ne T_{c||}$, as discussed in Sec.
\ref{sec-LRO_anisotropic_scattering}.
In the opposite limit, $d\ll\xi_s\ll\xi_0$, the aerogel medium is on average isotropic on length scales
larger than $\xi_s$. As a result the orbital p-wave components are unstable at the same transition
temperature, \ie there is a single $T_c$.
However, the transition and the relative stability of the possible phases will generally be 
modified by the short-range anisotropic scattering.
In both limits \emph{random anisotropy} leads to formation of orbital domains (LIM effect) on length
scales that are typically longer than either the aerogel or pair correlation lengths, $\xi_s$ and
$\xi_0$.
For \He\ in low-density silica aerogels, $\xi_s\lesssim\xi_0$, with the two correlation lengths being
comparable at high pressures, $p\gtrsim 15\,\mbox{bar}$. And as I discuss in Sec.
\ref{sec-LIM_anisotropic_scattering} the LIM effect is controlled not only by competition between orbital
order and orientational energetics at the scale of $\xi_s$, but also by random anisotropy in ballistic
transport. The latter determines the random field for Cooper pairs when $\xi_s\lesssim\xi_0$.

\subsection{Long-range order of Anisotropic Impurities} \label{sec-LRO_anisotropic_scattering}

Silica aerogels with exceptional homogeneity and global anisotropy have been fabricated,\cite{pol07}
and global anisotropy can be induced by uniaxial compression of an isotropic aerogel. 
Global anisotropy in these aerogels corresponds to long-range order of the locally anisotropic
scattering medium,
\be\label{strand_correlations_long-range}
\langle\hvs(\vr)\cdot\hvs(0)\rangle 
\xrightarrow[r\gg\xi_s]{}
s^2
\,,
\ee
where $0 < s^2 < 1$ measures the degree of long-range orientational order along
the uniaxial anisotropy, or ``strain'', axis, $\hvz$. Random fluctuations of the
anisotropy direction remain, but are uncorrelated over distances larger than $\xi_s$.

Long-range order of the strands is directly observable in transport properties of
normal \He\ in a globally anisotropic aerogel. At temperatures below the 
cross-over scale $T^{\star}\approx 20-30\,\mbox{mK}$ the transport of entropy and magnetization 
are determined by elastic scattering of quasiparticles from the aerogel structure. For example, the
thermal conductivity becomes anisotropic below $T^{\star}$,\cite{sau10}
\be
\kappa_{ij}
=
\frac{2\pi^2}{9}\,N_f\,(v_f T)\,\bar\ell_{ij}
\,,
\ee
where $\bar\ell_{ij}$ is the  transport \mfp\ tensor obtained from the Boltzmann-Landau transport 
equation with the collision integral determined by the elastic scattering rate in Eq. \ref{strand_model},
\be
\bar\ell_{ij}  = \bar\ell_{\perp}\left(\delta_{ij}-\hvz_{i}\hvz_{j}\right) 
           + \bar\ell_{\parallel}\,\hvz_{i}\hvz_{j}
\,,
\ee
where $\bar\ell_{\perp}=v_f\tau_{\perp}$ ($\bar\ell_{\parallel}=v_f\tau_{\parallel}$) is the transport
\mfp\ for heat transport perpendicular (parallel) to the anisotropy axis. Hydrodynamic transport averages
over length scales long compared to $\xi_s$, thus long-range orientational 
order determines the anisotropy
of the transport coefficients with,
\ber\label{tau_perp}
\frac{1}{\tau_{\perp}}      &=& \frac{1}{\bar\tau} + \onethird\Delta
% \\
% \label{tau_parallel}
\,,\quad\frac{1}{\tau_{\parallel}} = \frac{1}{\bar\tau} - \twothirds\Delta
% \\
% \label{tau_parallel}
% \frac{1}{\tau_{\parallel}} &=& \frac{1}{\bar\tau} - \twothirds\Delta
\\
\label{tau_average}
\frac{1}{\bar\tau} &=& w_0 
                 - \onethird\left(\twothirds\,w_{\perp}+\onethird\,w_{\parallel}\right)
\\
\label{tau_splitting}
\Delta              &=& 
                 - \onethird\,s^2\,\left(w_{\perp} - w_{\parallel}\right)
\,.
\eer
Note that the anisotropy in the scattering rates, $\Delta$, scales as the product of the long-range
orientational order of the anisotropic ``impurities'', $\sim s^2$, and the (local) anisotropy of the
p-wave scattering rates, $\sim(w_{\perp}-w_{\parallel})$. In the absence of orientational fluctuations
($s=1$) we obtain the maximal anisotropy in the scattering rates: 
$1/\tau_{\perp}=1/\tau_0 -\onethird w_{\perp}$, 
$1/\tau_{\parallel}=1/\tau_0 -\onethird w_{\parallel}$ and
$(1/\tau_{\perp}-1/\tau_{\parallel})=-\onethird(w_{\perp}-w_{\parallel})$.
In the absence of long-range orientational order ($s=0$), $\Delta =0$, and the local 
anisotropy gives an average isotropic scattering rate and \mfp, $\bar\ell=v_f\bar\tau$.

The \mfp\ anisotropy obtained from Eqs. \ref{tau_perp}-\ref{tau_splitting} determines the splitting of
the superfluid transition when the aerogel correlation length $\xi_s$ is smaller than the size of Cooper
pairs. This is the \emph{homogeneous scattering limit} in which quasiparticle scattering from the aerogel
is the dominant pair-breaking effect.\cite{thu96,thu98}
For a globally anisotropic scattering medium the pair-breaking effect is also symmetry breaking - 
lifting the degeneracy of the 3D p-wave orbital states and splitting the transition for pairing into 2D
orbital states $(\hvp_x,\hvp_y)$, and pairing into the 1D polar state, $\hvp_z$.\cite{thu96,aoy06}
The corresponding second-order GL coefficients that enter 
Eq. \ref{GLfunctional} for 2D and 1D orbital states are\cite{thu96}
\be\label{alpha_perp-||}
\bar\alpha_{\perp,\parallel}
=
\onethird\, N_f\,\Big[\ln(T/T_{c}) - \cS_{1}(x_{\perp,\parallel}\,T_{c}/T)\Big]
\,,
\ee
where $x_{\perp,\parallel} = \xi_0/\bar\ell_{\perp,\parallel}$ are the anisotropic pairbreaking
parameters, and $\cS_{1}(z)$ is the digamma function,
\be
\cS_{1}(z) 
= \sum_{n=0}^{\infty}
  \Big[\frac{1}{n+\tinyonehalf +\tinyonehalf z} 
- 
       \frac{1}{n+\tinyonehalf}
  \Big]
\,.
\ee
The instability temperatures for superfluid \He\ in a globally anisotropic medium are given by
$\alpha(T_{c_{\perp,\parallel}})=0$, and are the solutions to the Abrikosov-Gorkov
equation,\cite{abr61}
\be\label{Tc-perp-||}
\ln(T_{c}/T_{c_{\perp,\parallel}}) = \cS_{1}(x_{\perp,\parallel}\,T_{c}/T_{c_{\perp,\parallel}})
\,.
\ee
In the linear pair-breaking regime, we obtain 
\be\label{Tc-aerogel}
T_{c_{\perp,\parallel}} 
\simeq  T_{c}\,\left(1 - \frac{\pi^2}{4}\,\,\frac{\xi_0}{\bar\ell_{\perp,\parallel}} \right)
\,,
\ee
where $\xi_0=\hbar v_f/2\pi\,k_{\text{B}}\,T_{c_0}$ ($T_{c_0}$) is the pair correlation length
(transition temperature) for pure \He.
Thus, for $\bar\ell_{\perp} > \bar\ell_{\parallel}$ the first instability will be to a 2D orbital state
with the 2D orbital order parameter $\vec{a} \ne 0$, while for $\bar\ell_{\perp} < \bar\ell_{\parallel}$
the first instability will be into the 1D polar state with $b\ne 0$.
The orbital representation that is realized at the first instability for a ``stretched'' or
mechanically compressed aerogel depends on the microscopic mechanism(s) for the development of global
anisotropy in the ballistic path length distribution for quasiparticles in anisotropic
aerogel.
The splitting of the transition,
\be\label{Tc-splitting}
\frac{T_{c_{\perp}}-T_{c_{\parallel}}}{T_{c}}
\simeq 
-\frac{\pi^2}{4}\,\xi_0\,
\left(\frac{1}{\bar\ell_{\perp}} - \frac{1}{\bar\ell_{\parallel}}\right)
\,,
\ee
is proportional to the difference in the mean scattering rates, which scales as
$(1/\tau_{\perp}-1/\tau_{\parallel})=-\onethird\,s^2\,(w_{\perp}-w_{\parallel})$. Thus, the 
splitting of the transition may be significantly
smaller than the suppression of the superfluid transition from that of pure \He, which depends on the
total scattering rate (Eq. \ref{tau_perp}).
This provides a natural explanation for the relatively large $T_c$-suppression and relatively small $T_c$-splitting
shown Fig. \ref{fig-phase_diagram},
which at $18\,\mbox{bar}$ is a suppression of $T_c - T_{c_{\perp}}\simeq 0.84\,\mbox{mK}$, and 
a splitting $T_{c_{\perp}}-T_{c_{\parallel}}$ less than $0.28\,\mbox{mK}$.
This leads us to a prediction for \He\ in the stretched aerogel.\cite{pol12}
If long-range order of anisotropic scattering centers is the mechanism for stabilizing the 2D chiral ABM
state in stretched aerogel, then measurements of heat transport in normal \He\ should show anisotropy
with $\kappa_{\perp}/\kappa_{\parallel}=\bar\ell_{\perp}/\bar\ell_{\parallel} > 1$. 
Furthermore, if the ESP-2 state is the signature of the biaxial phase associated with the onset of the
polar distortion and a non-vanishing $0<T_{c_{\parallel}}<T_{c_{\perp}}$, then the ratio for
$\kappa_{\perp}/\kappa_{\parallel}$ can be predicted from the splitting of the transitions and the
$\beta$ parameters obtained from NMR and thermodynamic measurements on these phases.

Finally, note that for sufficiently strong anisotropy the second instability may be suppressed to zero,
or to temperatures well outside the GL limit.
This limit is likely relevant to superfluid \He\ reported in the
newly discovered ``nematic'' aerogels with \mfp's of $\bar\ell_{\parallel}=850\,\mbox{nm}$ and
$\bar\ell_{\perp}=450\,\mbox{nm}$ based on measurements of spin diffusion in the normal
state.\cite{ask12}
The anisotropy ratio, $\bar\ell_{\parallel}/\bar\ell_{\perp}\approx 2$, favors a transition from the
normal state into the \emph{1D polar phase} based on this theory of homogeneous anisotropic pairbreaking.
Nematic aerogels appear to be inhomogeneous on the scale of $\xi_0$, so this theory may account
for a transition to the polar state near $T_{c_{\parallel}}$, where $\xi(T)\gg\xi_a$, but is likely
outside the regime of validity to describe the low temperature phases.

\subsection{Random Anisotropy}\label{sec-LIM_anisotropic_scattering}

%----------- LIM Domain Size vs. Pressure -------------------
\begin{figure}[!]
\includegraphics[width=0.995\linewidth,keepaspectratio]{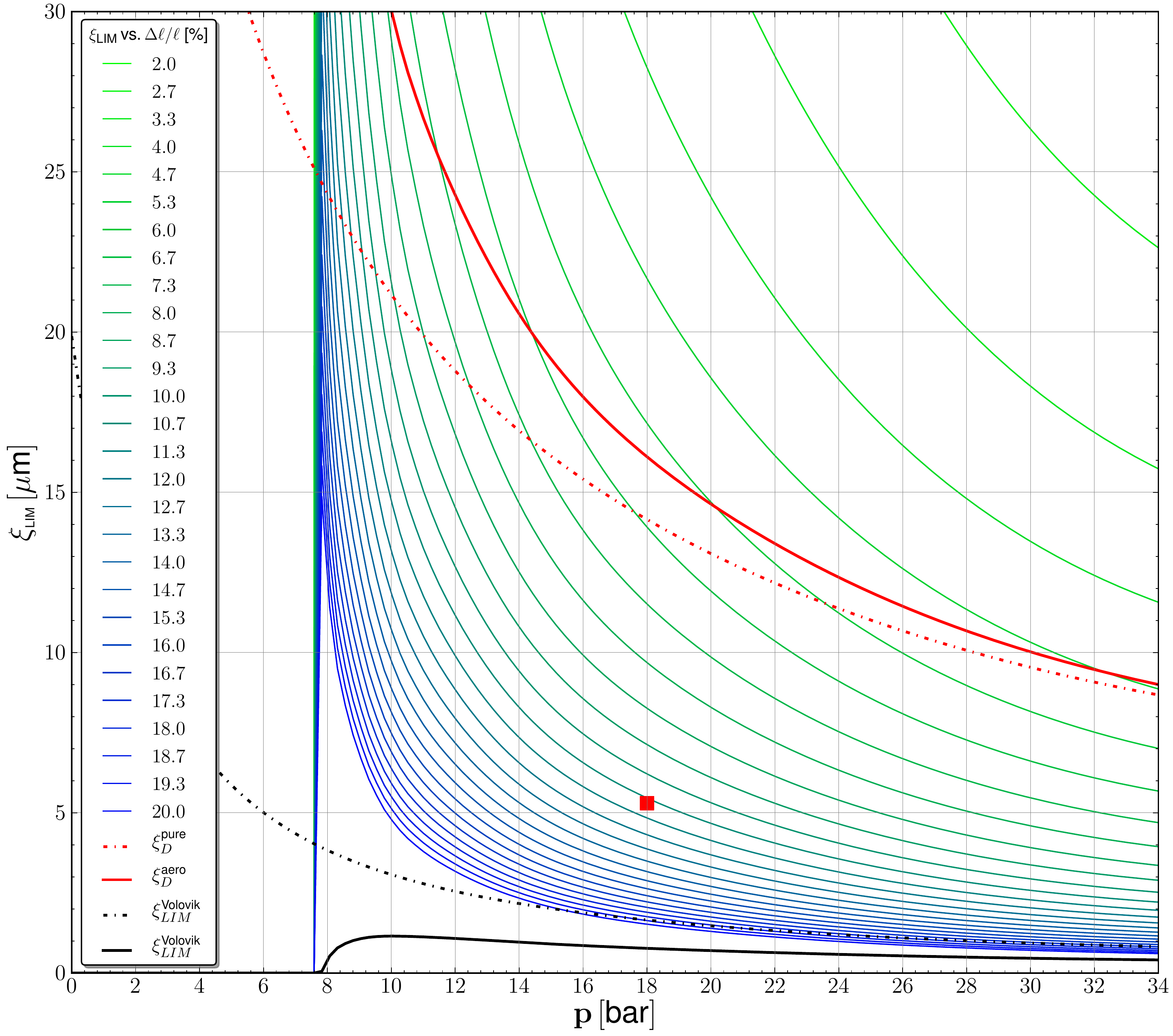}
\caption{\small
Larkin-Imry-Ma domain size for the biaxial orbital order parameter as a function of pressure for aerogel
with average \emph{mfp}, $\ell=120\,\mbox{nm}$, strand correlation length, $\xi_{s}=20\,\mbox{nm}$
and anisotropy ranging from $\Delta{\ell}/{\ell} = 2\,\% - 20\,\%$ [\blue{blue}-\darkgreen{green}
curves].
Volovik's result for the LIM length [{\bf black dashed} curve], while the [{\bf black solid} curve]
includes the renormalization of the parameters defining the LIM length due to pairbreaking 
for an aerogel with $T_{c}(p)$ determined by $\bar{x}=\xi_0/\bar\ell$.
The dipole coherence length, $\xi_{\text{D}}$, is shown as the [\red{solid red} curve] for the same
\emph{mfp}, and for pure \He\ as the [\red{dashed red} curve].
The point [\red{\tiny$\blacksquare$}] at $18\,\mbox{bar}$ is $\xi_{\text{LIM}}\approx 5.3\,\mu\mbox{m}$
determined from the width of the negative jump in the NMR shift shown in Fig. \ref{fig-NMR_shifts}.
\label{fig-LIM_size}
}
\end{figure}
%-------------------------------------------------------------

The biaxial phase described in Secs. \ref{sec-biaxial_phase} and \ref{sec-NMR_ESP-2} that onsets below
$T_{c_2}$ has a continuous degeneracy corresponding to rotations of the orbital triad
$\{\hvm,\hvn,\hvl\}$ about the uniaxial ``stretch'' axis, $\hvz$. The latter is interpreted here as a
manifestation of long-range orientational order of the scattering centers with $\langle\hvs\rangle = s
\hvz$.
Fluctuations in the local anisotropy axis, $\vs'(\vr) = \hvs - s\hvz$, couple to the 
orbital order parameter and partially destroy the long-range orientational order 
of the biaxial phase.

The \emph{random} anisotropy energy is defined by the fluctuations in the pair-breaking effect
given by
\be\label{random_field_energy}
\Delta\Omega_{\text{an}} = \int_{\text{V}}\,d^3\,\delta\alpha_{ij}(\vr)\,\cA_{i}\,\cA_{j}^{*}
\,,
\ee
where the leading order term of $\delta\alpha_{ij}(\vr)=\alpha_{ij}(\vr)-{\bar\alpha}_{ij}$
for weak global anisotropy ($s^2 \ll 1$) is
\ber
\delta\alpha_{ij} &=& -\onehalf\,\delta\alpha\,\left(\vs'_{i}(\vr)\,\vs'_{j}(\vr) - \onethird\right)
\\
\delta\alpha &=& (\pi^2/4)\,N_f\,\xi_{0}\,(1/\ell_{\parallel}-1/\ell_{\perp})\,
                 \left[\cS_2(\bar{x})/\cS_2(0)\right]
\,.\qquad
\label{eq-random_field_strength}
\eer
Note that the random anisotropy is enhanced compared to the global anisotropy by the factor $1/s^2$, i.e.
$(1/\ell_{\parallel}-1/\ell_{\perp})=\onethird(w_{\parallel}-w_{\perp})$ $=$
$(1/\bar\ell_{\parallel}-1/\bar\ell_{\perp})/s^2$.
Also I include the mean-field pair-breaking effects with 
$\bar{x}=\hbar/2\pi T_c\bar\tau = \xi_0/\bar\ell$, the 
suppression of $T_c$ by impurity scattering and the functions,
\ber
\cS_{p}(x) &\equiv& \sum_{n\ge 0}\left(n+\onehalf +\onehalf x\right)^{-p}\,,\quad p > 1
\,,
\eer
that renormalize the GL coefficients.\cite{thu96,sau03}

The random anisotropy energy obtained from Eqs. \ref{random_field_energy}-\ref{eq-random_field_strength}
and \ref{OP_tensor_real3}, the completeness relation,
$\delta_{ij}=\hvm_{i}\hvm_{j}+\hvn_{i}\hvn_{j}+\hvl_{i}\hvl_{j}$ and 
$\Delta^2_z\ll\Delta^2_{s,d}$ for $T\lesssim T_{c_2}$ becomes,
\be\label{random_field_energy2}
\Delta\Omega_{\text{an}} 
= 
-\onehalf\,\delta\alpha\,\Delta_{s}^2\,
 \int_{\text{V}}\,d^3r\,\left[(\vs'(\vr)\cdot\hvl(\vr))^2 - \onethird\right]
\,,
\ee
where $\hvl(\vr)$ is the local chiral axis in the biaxial phase. The mean field orientation for $\hvl$ is
fixed on the cone shown in Fig. \ref{fig-biaxial_NMR_geometry}. The transverse orbital order parameter,
$\delta\hvl=\hvl - \bar{l}_{z}\hvz$, is degenerate and can point in any direction in the base of
the cone.

The fluctuations in the anisotropy of the scattering medium are correlated on the scale of $\xi_s$. Thus,
the minimum of the random anisotropy energy is achieved by having the chiral axis $\hvl(\vr)$ 
``track'' the 
transverse fluctuations in anisotropy, i.e. $\delta\hvl(\vr) \parallel \vs'_{\perp}(\vr)$. 
However, tracking the anisotropy on the scale of $\xi_s$ costs gradient energy of order,
\be\label{gradient_energy}
\Delta\Omega_{\text{grad}} 
=
\kappa\,\Delta_{s}^2\,\int_{\text{V}}d^3r\,\vert\grad_i\,\hvl_j\vert^2
\approx
V\,\left(\,\kappa\,\sin^2\vartheta\,\Delta_{s}^2\,\xi_{s}^{-2}\right)
\,,
\ee
with $\kappa = \fourfifteenths(7\zeta(3)/8) N_f \xi_0^2\,[\cS_{3}(\bar{x})/\cS(0)]$,\cite{thu96}
which gives a gradient energy that is larger than the condensation energy for $\xi_{s}<\xi_{0}$
and $T< T_{c_2}$.
The balance between the local anisotropy energy and the gradient energy leads to the
\emph{partial} destruction of long-range orbital order for the biaxial phase in which
$\langle\hvl\rangle=\bar{l}_{z}\,\hvz$,
but long-range order of the transverse orbital order parameter, 
$\delta\hvl$, is destroyed.

The competition between the random anisotropy energy (Eq. \ref{random_field_energy2}) and
the gradient energy leads to short-range transverse orbital order over  
length scales, $\xi_{\text{LIM}}\gg\xi_s$, i.e. the Larkin-Imry-Ma (LIM) domain size.
The argument here is similar to that discussed by Volovik,\cite{vol08} but with different 
energy and length scales ultimately determining $\xi_{\text{LIM}}$.
In the presence of the random anisotropy field, the biaxial phase can avoid large gradient energies by
allowing $\delta\hvl$ to remain nearly uniform over length scales of order 
$\xi_{\text{LIM}}\gg\xi_{0}>\xi_{s}$. Thus, $\xi_{\text{LIM}}$ is the
domain size characterizing the short-range transverse orbital order.
The cost in gradient energy to bend the order parameter over the same length
scale is significantly reduced compared to Eq. \ref{gradient_energy}, but so too
is the gain in the random anisotropy energy. 
The latter is reduced by the fraction of anisotropy domains with $\vs'$ 
favorably aligned with the transverse order parameter, $\delta\hvl$, within an orbital domain.
Within an orbital domain of volume $V_{\text{LIM}}=\xi_{\text{LIM}}^3$ the mean number of domains of the
anisotropy axis $\vs'_{\perp}$ is $N_{s} = (\xi_{\text{LIM}}/\xi_{s})^3 \gg 1$.
The fraction of anisotropy domains that can favorably be aligned is the fluctuation ratio,
$
f_{\text{an}} = \Delta N_{s}^{\text{rms}}/N_{s} = 1/\sqrt{N_{s}}\approx (\xi_s/\xi_{\text{LIM}})^{3/2}
$.
With this estimate the optimal domain size is determined by minimizing the fluctuation of the anisotropy
energy together with the gradient energy,\cite{imr75,vol08}
\be
\Delta\Omega_{\text{fluc}} = 
V\,\sin^2\vartheta\,\Delta_{s}^2
\left(
-\onethird\,\delta\alpha\,\,(\xi_{s}/\xi_{\text{LIM}})^{3/2}
+
\kappa\,\xi_{\text{LIM}}^{-2}
\right)
\,,
\ee
which gives the LIM domain size 
\ber\label{LIM_size}
\hspace*{-10mm}
\xi_{\text{LIM}} 
&=&\left(\frac{4\kappa}{\delta\alpha}\right)^2\,\xi_{s}^{-3}
% \\
= C_{\text{L}}\,\frac{\xi_{0}^2\xi_{s}^{-3}}{[1/\ell_{\parallel}-1/\ell_{\perp}]^2}
\,,
\\
\hspace*{-7mm}
C_{\text{L}}(\bar{x}) &=& \left(\frac{4}{15}\right)^2
                          \left(\frac{4}{\pi}\right)^4
                          \left(\frac{7\zeta(3)}{8}\right)^2\,
                          \left[\frac{\cS_{3}(\bar{x})/\cS_{3}(0)}
                               {\cS_{2}(\bar{x})/\cS_{2}(0)}\right]^2
\label{Impurity_renormalization}
\,,
\eer
where $C_{\text{L}}(\bar{x})$ includes the renormalization of $\kappa$ and $\delta\alpha$
due to the breaking of Cooper pairs by quasiparticle scattering, parametrized by
$\bar{x}=\xi_{0}/\bar{\ell}$.
In the limit $\bar{x}\rightarrow 0$, $C_{\text{L}}(0)\simeq 0.21$. 

Figure \ref{fig-LIM_size} shows the pressure dependence of the LIM domain size calculated from Eqs.
\ref{LIM_size}-\ref{Impurity_renormalization} for an anisotropic aerogel defined by an average \mfp,
$\bar\ell = 120\,\mbox{nm}$ and the random field anisotropy of the aerogel expressed in terms of the
anisotropy in the \mfp's, $\delta\ell/\ell = 2-20\,\%$.\footnote{The pressure dependent Fermi-liquid
parameters of \He\ are taken from the ``Helium Calculator'' [url:
http://spindry.phys.northwestern.edu/he3.htm].} The mean-field effect of impurity scattering is included
via the impurity scattering renormalization of the transition temperature, $T_c/T_{c_0}$, the gradient
coefficient, $\kappa\sim\cS_3(\bar{x})/\cS_3(0)$, and the random anisotropy coefficient,
$\delta\alpha\sim \cS_2(\bar{x})/\cS_2(0)$.
I also show the pressure dependence of the dipole coherence length. The LIM effect on the NMR frequency
shift depends on the relative size of the orbital domains to the dipole coherence length,
$\xi_{\text{D}}=\sqrt{\kappa/g_{\text{D}}}$, since $\vec\vd$ can adjust to the local orbital order only
on length scales larger than $\xi_{\text{D}}$.\cite{vol08}
The dipole coupling constant is un-renormalized by impurity scattering,\cite{thu96} and is fixed at each
pressure by the measured bulk A-phase NMR shift.\cite{thu87} The resulting curves for $\xi_{\text{D}}$
for bulk \He\ (dotted red curve) and \Heaero\ for $\bar\ell=120\,\mbox{nm}$ (solid red curve) are shown
for comparison with $\xi_{\text{LIM}}$. Note that the critical pressure for this aerogel, below which
superfluidity is suppressed is $p_{c}\approx 8\,\mbox{bar}$.
The main result is that the orbital domain size is typically $\xi_{\text{LIM}}\lesssim\xi_{\text{D}}$
over the full pressure range, and may be larger than $\xi_{\text{D}}$ for relatively weak random
anisotropy, $\Delta\ell/\ell \lesssim 5\%$.

Also shown in Fig. \ref{fig-LIM_size} is the LIM domain size obtained from the rigid cylinder 
model of random anisotropy proposed by Volovik - the solid black curve includes the impurity 
renormalization of the gradient coefficient (not shown in Eq. \ref{LIM-Volovik}).
The suppression of $\xi_{\text{LIM}}^{\text{Volovik}}\rightarrow 0$ for $p\rightarrow p_{c}$ is likely an
artifact of the single-impurity model for the random anisotropy energy from Rainer and
Vuorio.\cite{rai77} Away from $p_c$ $\xi_{\text{LIM}}^{\text{Volovik}} < 1\mu\mbox{m}$ over the full
pressure range, implying
that orbital order is destroyed by a relatively strong random anisotropy field over length scales much
smaller than $\xi_{\text{D}}$ at all pressures ($\xi_{\text{LIM}}^{\text{Volovik}}\approx 0.77\mu\mbox{m}$ compared to $\xi{\text{D}}\approx 16.1\,\mu\mbox{m}$ at $18\,\mbox{bar}$).
The conclusion here is that the random anisotropy field that is relevant to the destruction of orbital
order, and the LIM domain size, can be much weaker, originating from mesoscale structures
that are much larger than atomic scale, $d\ll \delta r \ll \xi_a$, and are responsible for local
anisotropy in the quasiparticle scattering rate.

In the weak anisotropy limit the LIM averaging of the dipole energy breaks down and we expect
a distribution of NMR shifts resulting from the distribution of spatial variations of the dipole energy
on the scale $\xi_{\text{LIM}}\sim\xi{\text{D}}$. 
The NMR spectra for the stretched aerogel - particularly the rapid reduction in the shift below
$T_{c_2}$ and the broadening of the line - suggests that $\xi_{\text{LIM}}\lesssim\xi_{\text{D}}$.
In this case the orienting effect on the orbital order parameter of the biaxial phase by the dipole
energy can be treated perturbatively.
For $T\ll T_{c_2}$ the polar distortion is established, $\hvl$ is oriented off the anisotropy axis and
the transverse orbital order parameter, $\delta\hvl=\sin\vartheta(\cos\varphi\hvx+\sin\varphi\hvy)$, is
degenerate on the cone in Fig. \ref{fig-biaxial_NMR_geometry}.
Optimizing the random anisotropy energy and the gradient energy for the transverse orbital order leads to
the optimal orbital domain size given by Eq. \ref{LIM_size}. These two energies are of the same order
with an overall magnitude that scales as $\Delta_{s}^2\sin^2\vartheta\sim\Delta_{z}^2\sim(1-T/T_{c_2})$.
Thus, sufficiently close to $T_{c_2}$ the dipole energy will become comparable to the optimized random
field domain-alignment energy. 
The dipole energy can now compete to align the transverse orbital order parameter and recover dipole
energy that was ``lost'' by averaging $\delta\hvl$ on the cone.
At high fields, $\omega_{\,\text{L}}\gg \Omega_{\text{A}}$, $\vec{\vd}$, is fixed perpendicular to the
field, e.g. $\vec\vd \perp \hvy$ in Fig. \ref{fig-biaxial_NMR_geometry}.
The competition between the fluctuation contribution to the dipole energy and random field
domain-alignment energies converts the jump in $\Delta\omega$ at $T_{c_2}$ into a cross-over with a
transition width, $\delta T_{c_2}$, given by the temperature at which the ``missing'' dipole energy
becomes equal to the stiffness of the biaxial LIM state,
\ber
\hspace*{-5mm}
\onehalf g_{\text{D}}\Delta_{\text{A}}^2 
= 
\kappa\,\sin^2\vartheta\,\Delta_{s}^2\,\xi_{\text{LIM}}^{-2}
\simeq 2\,\kappa\,\Delta_{z}(T_{c_2}+\delta T_{c_2})^2\,\xi_{\text{LIM}}^{-2}
\\
\leadsto\quad
\delta T_{c_2} 
= T_{c_2}\,\times\,\onefourth\,
                 \left(\frac{\Delta_{\text{A}}}{\bar\Delta_{z}}\right)^2\,
                 \left(\frac{\xi_{\text{LIM}}}{\xi_{\text{D}}}\right)^2
\,.
\hspace*{5mm}
\eer
For the $\bar\Delta_{z}/\Delta_{\text{A}}\approx 2$ 
at $T_{c_2}$ we obtain the estimate of $\xi_{\text{LIM}}\approx 5\mu\mbox{m}$ from the 
observed transition width of $\delta T_{c_2}/T_{c_2}\approx 0.1$ at $18\,\mbox{bar}$,
which based on random field anisotropy in the scattering rate corresponds to 
$\Delta\ell/\ell \simeq 11\%$.
For comparison the authors of Ref. \onlinecite{pol12} characterize the ``stretched'' aerogel in their
NMR experiments by uniaxial strain $\varepsilon_{zz}\simeq 14\%$.

\subsection*{Summary and Conclusions}

The discovery of equal-spin-pairing (ESP) phases of superfluid \He\ in highly porous \emph{anisotropic}
silica aerogels provides us with a unique condensed matter system for investigation of 
remarkable phases that may be realized in systems with broken continuous symmetries and competing
effects from disorder.
A Ginzburg-Landau (GL) theory for orbital p-wave phases in a medium with both global and random
anisotropy was developed. Global anisotropy gives rise to multiple ordered phases that are characterized
as 2D (chiral or in-plane polar), 1D axial aligned polar states and a ``mixed'' symmetry phase that
exhibit both biaxial and chiral order, depending on the nature of the \emph{global} anisotropy, e.g.
stretched vs. compressed anisotropy.
The 2D chiral phase is an ABM state with the chiral axis aligned \underline{along} the anisotropy axis in 
the case of ``stretched'' aerogels, i.e. $\hvl=\pm\hvz$. 
%
% Analysis of the NMR signatures shows that this state is the order parameter for the ESP-1 phase 
% of the experiments reported by Pollanen et al.\cite{pol12}.
%
The NMR signatures of the 2D chiral phase with $\hvl\parallel\pm\hvz$ are in quantitative agreement with
the ESP-1 phase of \He\ in ``stretched aerogel'' reported in Ref. \cite{pol12}, including a large
transverse shift, the tipping angle dependence \underline{and} a narrow linewidth. The ESP-1 phase can
not be identified with a chiral phase with $\hvl\perp\hvz$. Not only is this phase excluded by symmetry
for a uniform uniaxial medium, if it were present as a second, low-temperature phase it would exhibit a
reduced NMR shift and a broadened NMR line due to the LIM effect.
Additional support for this identification is provided by recent NMR measurements\cite{li13} on \He\
infused into the same ``stretched aerogel'' as in Ref. \cite{pol12}, but with the static NMR field along
the strain axis and the \emph{rf} field perpendicular to the strain axis. In this orientation the NMR
shift that onsets at the same $T_{c_1}$ as in Ref. \cite{pol12} is \emph{negative} as is expected for an
ESP-1 phase with $\hvl\parallel\pm\hvz$. In contrast a chiral phase with $\hvl\perp\hvz$ should exhibit a
positive shift for both field configurations. Furthermore, the results of Li et al. exclude the normal to
1D polar phase scenario, since the polar phase, were it the present, would show a large positive shift
onsetting at a temperature \emph{above} $T_{c_1}$ as measured by Pollanen et al. No such transition is
observed. Thus, the analysis presented here combined with the experiments of Refs. \cite{pol12} and
\cite{li13} show that the ESP-1 phase is a chiral ABM state with $\hvl\pm\hvz$.

The biaxial phase spontaneously breaks the rotational symmetry about the global anisotropy axis,
and is identified with the ESP-2 phase of \He\ in stretched aerogel. 
This identification depends upon the interplay between the continuous degeneracy of the biaxial phase,
associated with broken $\gaugeorbit$ rotational symmetry, and \emph{random} anisotropy associated with
the structure of the aerogel.
Comparison of the NMR spectrum for the ESP-2 phase with theoretical predictions for the NMR frequency
shifts provides strong evidence for identifying the ESP-2 as the biaxial state, partially
\emph{disordered} by random anisotropy.
The analysis is based on an expansion of Volovik's original random field model for \Heaero. I argue that
the random anisotropy field results from mesoscopic structures in silica aerogels - coarse-grained on the
atomic scale - and formulated in terms of local anisotropy in the scattering of quasiparticles in an
aerogel with orientational correlations.
Long-range order of locally anisotropic scattering centers is responsible for the splitting of the
transition for 1D and 2D orbital states.

Further tests of this theoretical description of anisotropic pairbreaking, random anisotropy, and the
stability of unique orbital phases of superfluid \He\ are possible with  
transport experiments on the same, or similarly prepared anisotropic aerogels.

\subsection*{Acknowledgements}

This research is supported by the National Science Foundation (Grants DMR-0805277 and DMR-1106315).
I thank Sarosh Ali, supported on this project, for developing the DLCA code used to generate simulation
images of the nanoscale structure of 98\% aerogel.
I acknowledge many discussions and explanations of experiments on \He\ in anisotropic aerogels with Bill
Halperin, Johannes Pollanen, Jia Li, Jeevak Parpia and Vladimir Dimitriev, as well as illuminating
discussions on this work with Vladimir Mineev.
I acknowledge the hospitality and support of the Aspen Center for Physics where part of this work
was carried out.

%---------------------- References ---------------------------------------
% \bibliographystyle{apsrev}
% \bibliography{QFS,Books,CM}

\begin{thebibliography}{44}
\expandafter\ifx\csname natexlab\endcsname\relax\def\natexlab#1{#1}\fi
\expandafter\ifx\csname bibnamefont\endcsname\relax
  \def\bibnamefont#1{#1}\fi
\expandafter\ifx\csname bibfnamefont\endcsname\relax
  \def\bibfnamefont#1{#1}\fi
\expandafter\ifx\csname citenamefont\endcsname\relax
  \def\citenamefont#1{#1}\fi
\expandafter\ifx\csname url\endcsname\relax
  \def\url#1{\texttt{#1}}\fi
\expandafter\ifx\csname urlprefix\endcsname\relax\def\urlprefix{URL }\fi
\providecommand{\bibinfo}[2]{#2}
\providecommand{\eprint}[2][]{\url{#2}}

\bibitem[{\citenamefont{Porto and Parpia}(1995)}]{por95}
\bibinfo{author}{\bibfnamefont{J.}~\bibnamefont{Porto}} \bibnamefont{and}
  \bibinfo{author}{\bibfnamefont{J.}~\bibnamefont{Parpia}},
  \bibinfo{journal}{Phys. Rev. Lett.} \textbf{\bibinfo{volume}{74}},
  \bibinfo{pages}{4667} (\bibinfo{year}{1995}).

\bibitem[{\citenamefont{Sprague et~al.}(1995)\citenamefont{Sprague, Haard,
  Kycia, Rand, Lee, Hamot, and Halperin}}]{spr95}
\bibinfo{author}{\bibfnamefont{D.}~\bibnamefont{Sprague}},
  \bibinfo{author}{\bibfnamefont{T.}~\bibnamefont{Haard}},
  \bibinfo{author}{\bibfnamefont{J.}~\bibnamefont{Kycia}},
  \bibinfo{author}{\bibfnamefont{M.}~\bibnamefont{Rand}},
  \bibinfo{author}{\bibfnamefont{Y.}~\bibnamefont{Lee}},
  \bibinfo{author}{\bibfnamefont{P.}~\bibnamefont{Hamot}}, \bibnamefont{and}
  \bibinfo{author}{\bibfnamefont{W.}~\bibnamefont{Halperin}},
  \bibinfo{journal}{Physical Review Letters} \textbf{\bibinfo{volume}{75}},
  \bibinfo{pages}{661} (\bibinfo{year}{1995}).

\bibitem[{\citenamefont{Pollanen et~al.}(2008)\citenamefont{Pollanen, Shirer,
  Blinstein, Davis, Choi, Lippman, Halperin, and Lurio}}]{pol08}
\bibinfo{author}{\bibfnamefont{J.}~\bibnamefont{Pollanen}},
  \bibinfo{author}{\bibfnamefont{K.~R.} \bibnamefont{Shirer}},
  \bibinfo{author}{\bibfnamefont{S.}~\bibnamefont{Blinstein}},
  \bibinfo{author}{\bibfnamefont{J.~P.} \bibnamefont{Davis}},
  \bibinfo{author}{\bibfnamefont{H.}~\bibnamefont{Choi}},
  \bibinfo{author}{\bibfnamefont{T.~M.} \bibnamefont{Lippman}},
  \bibinfo{author}{\bibfnamefont{W.~P.} \bibnamefont{Halperin}},
  \bibnamefont{and} \bibinfo{author}{\bibfnamefont{L.~B.} \bibnamefont{Lurio}},
  \bibinfo{journal}{Journal Of Non-Crystalline Solids}
  \textbf{\bibinfo{volume}{354}}, \bibinfo{pages}{4668} (\bibinfo{year}{2008}).

\bibitem[{\citenamefont{Leggett}(1975)}]{leg75}
\bibinfo{author}{\bibfnamefont{A.~J.} \bibnamefont{Leggett}},
  \bibinfo{journal}{Rev. Mod. Phys.} \textbf{\bibinfo{volume}{47}},
  \bibinfo{pages}{331} (\bibinfo{year}{1975}).

\bibitem[{\citenamefont{Pollanen et~al.}(2011)\citenamefont{Pollanen, Li,
  Collett, Gannon, and Halperin}}]{pol11}
\bibinfo{author}{\bibfnamefont{J.}~\bibnamefont{Pollanen}},
  \bibinfo{author}{\bibfnamefont{J.~I.~A.} \bibnamefont{Li}},
  \bibinfo{author}{\bibfnamefont{C.~A.} \bibnamefont{Collett}},
  \bibinfo{author}{\bibfnamefont{W.~J.} \bibnamefont{Gannon}},
  \bibnamefont{and} \bibinfo{author}{\bibfnamefont{W.~P.}
  \bibnamefont{Halperin}}, \bibinfo{journal}{Phys. Rev. Lett.}
  \textbf{\bibinfo{volume}{107}}, \bibinfo{pages}{195301}
  (\bibinfo{year}{2011}).

\bibitem[{\citenamefont{Balian and Werthamer}(1963)}]{bal63}
\bibinfo{author}{\bibfnamefont{R.}~\bibnamefont{Balian}} \bibnamefont{and}
  \bibinfo{author}{\bibfnamefont{N.~R.} \bibnamefont{Werthamer}},
  \bibinfo{journal}{Phys. Rev.} \textbf{\bibinfo{volume}{131}},
  \bibinfo{pages}{1553} (\bibinfo{year}{1963}).

\bibitem[{\citenamefont{Choi et~al.}(2004)\citenamefont{Choi, Yawata, Haard,
  Davis, Gervais, Mulders, Sharma, Sauls, and Halperin}}]{cho04a}
\bibinfo{author}{\bibfnamefont{H.}~\bibnamefont{Choi}},
  \bibinfo{author}{\bibfnamefont{K.}~\bibnamefont{Yawata}},
  \bibinfo{author}{\bibfnamefont{T.}~\bibnamefont{Haard}},
  \bibinfo{author}{\bibfnamefont{J.}~\bibnamefont{Davis}},
  \bibinfo{author}{\bibfnamefont{G.}~\bibnamefont{Gervais}},
  \bibinfo{author}{\bibfnamefont{N.}~\bibnamefont{Mulders}},
  \bibinfo{author}{\bibfnamefont{P.}~\bibnamefont{Sharma}},
  \bibinfo{author}{\bibfnamefont{J.}~\bibnamefont{Sauls}}, \bibnamefont{and}
  \bibinfo{author}{\bibfnamefont{W.}~\bibnamefont{Halperin}},
  \bibinfo{journal}{Phys. Rev. Lett.} \textbf{\bibinfo{volume}{93}},
  \bibinfo{pages}{145301} (\bibinfo{year}{2004}).

\bibitem[{\citenamefont{Fisher et~al.}(2003)\citenamefont{Fisher, Guenault,
  Mulders, and Pickett}}]{fis03}
\bibinfo{author}{\bibfnamefont{S.}~\bibnamefont{Fisher}},
  \bibinfo{author}{\bibfnamefont{A.}~\bibnamefont{Guenault}},
  \bibinfo{author}{\bibfnamefont{N.}~\bibnamefont{Mulders}}, \bibnamefont{and}
  \bibinfo{author}{\bibfnamefont{G.}~\bibnamefont{Pickett}},
  \bibinfo{journal}{Phys. Rev. Lett.} \textbf{\bibinfo{volume}{91}},
  \bibinfo{pages}{105303} (\bibinfo{year}{2003}).

\bibitem[{\citenamefont{Sauls et~al.}(2005)\citenamefont{Sauls, Bunkov, Collin,
  Godfrin, and Sharma}}]{sau05}
\bibinfo{author}{\bibfnamefont{J.~A.} \bibnamefont{Sauls}},
  \bibinfo{author}{\bibfnamefont{Y.~M.} \bibnamefont{Bunkov}},
  \bibinfo{author}{\bibfnamefont{E.}~\bibnamefont{Collin}},
  \bibinfo{author}{\bibfnamefont{H.}~\bibnamefont{Godfrin}}, \bibnamefont{and}
  \bibinfo{author}{\bibfnamefont{P.}~\bibnamefont{Sharma}},
  \bibinfo{journal}{Phys. Rev. B} \textbf{\bibinfo{volume}{72}},
  \bibinfo{pages}{024507} (\bibinfo{year}{2005}).

\bibitem[{\citenamefont{Sharma and Sauls}(2001)}]{sha01}
\bibinfo{author}{\bibfnamefont{P.}~\bibnamefont{Sharma}} \bibnamefont{and}
  \bibinfo{author}{\bibfnamefont{J.~A.} \bibnamefont{Sauls}},
  \bibinfo{journal}{J. Low Temp. Phys.} \textbf{\bibinfo{volume}{125}},
  \bibinfo{pages}{115} (\bibinfo{year}{2001}).

\bibitem[{\citenamefont{Thuneberg et~al.}(1996)
          \citenamefont{Thuneberg, Yip, Fogelstr\"om, and Sauls}}]{thu96}
\bibinfo{author}{\bibfnamefont{E.~V.} \bibnamefont{Thuneberg}},
  \bibinfo{author}{\bibfnamefont{S.-K.} \bibnamefont{Yip}},
  \bibinfo{author}{\bibfnamefont{M.}~\bibnamefont{Fogelstr\"om}},
  \bibnamefont{and} \bibinfo{author}{\bibfnamefont{J.~A.} \bibnamefont{Sauls}},
  \bibinfo{journal}{arXiv} 
  \textbf{\bibinfo{volume}{\href{http://arxiv.org/abs/cond-mat/9601148v1}{cond-mat/9601148v1}}},
  \bibinfo{pages}{4} (\bibinfo{year}{1996}).

\bibitem[{\citenamefont{Thuneberg et~al.}(1998)\citenamefont{Thuneberg, Yip,
  Fogelstr\"om, and Sauls}}]{thu98}
\bibinfo{author}{\bibfnamefont{E.~V.} \bibnamefont{Thuneberg}},
  \bibinfo{author}{\bibfnamefont{S.-K.} \bibnamefont{Yip}},
  \bibinfo{author}{\bibfnamefont{M.}~\bibnamefont{Fogelstr\"om}},
  \bibnamefont{and} \bibinfo{author}{\bibfnamefont{J.~A.} \bibnamefont{Sauls}},
  \bibinfo{journal}{Phys. Rev. Lett.} \textbf{\bibinfo{volume}{80}},
  \bibinfo{pages}{2861} (\bibinfo{year}{1998}).

\bibitem[{\citenamefont{Aoyama and Ikeda}(2006)}]{aoy06}
\bibinfo{author}{\bibfnamefont{K.}~\bibnamefont{Aoyama}} \bibnamefont{and}
  \bibinfo{author}{\bibfnamefont{R.}~\bibnamefont{Ikeda}},
  \bibinfo{journal}{Phys Rev. B} \textbf{\bibinfo{volume}{73}}
  (\bibinfo{year}{2006}).

\bibitem[{\citenamefont{Vicente et~al.}(2005)\citenamefont{Vicente, Choi, Xia,
  Halperin, Mulders, and Lee}}]{vic05}
\bibinfo{author}{\bibfnamefont{C.~L.} \bibnamefont{Vicente}},
  \bibinfo{author}{\bibfnamefont{H.~C.} \bibnamefont{Choi}},
  \bibinfo{author}{\bibfnamefont{J.~S.} \bibnamefont{Xia}},
  \bibinfo{author}{\bibfnamefont{W.~P.} \bibnamefont{Halperin}},
  \bibinfo{author}{\bibfnamefont{N.}~\bibnamefont{Mulders}}, \bibnamefont{and}
  \bibinfo{author}{\bibfnamefont{Y.}~\bibnamefont{Lee}},
  \bibinfo{journal}{Phys. Rev. B} \textbf{\bibinfo{volume}{72}},
  \bibinfo{pages}{094519} (\bibinfo{year}{2005}).

\bibitem[{\citenamefont{Bennett et~al.}(2011)\citenamefont{Bennett, Zhelev,
  Smith, Pollanen, Halperin, and Parpia}}]{ben11}
\bibinfo{author}{\bibfnamefont{R.}~\bibnamefont{Bennett}},
  \bibinfo{author}{\bibfnamefont{N.}~\bibnamefont{Zhelev}},
  \bibinfo{author}{\bibfnamefont{E.}~\bibnamefont{Smith}},
  \bibinfo{author}{\bibfnamefont{J.}~\bibnamefont{Pollanen}},
  \bibinfo{author}{\bibfnamefont{W.}~\bibnamefont{Halperin}}, \bibnamefont{and}
  \bibinfo{author}{\bibfnamefont{J.}~\bibnamefont{Parpia}},
  \bibinfo{journal}{Phys. Rev. Lett.} \textbf{\bibinfo{volume}{107}}
  (\bibinfo{year}{2011}).

\bibitem[{\citenamefont{Anderson and Morel}(1960)}]{and60}
\bibinfo{author}{\bibfnamefont{P.~W.} \bibnamefont{Anderson}} \bibnamefont{and}
  \bibinfo{author}{\bibfnamefont{P.}~\bibnamefont{Morel}},
  \bibinfo{journal}{Phys. Rev. Lett.} \textbf{\bibinfo{volume}{5}},
  \bibinfo{pages}{136} (\bibinfo{year}{1960}).

\bibitem[{\citenamefont{Brinkman and Anderson}(1973)}]{bri73a}
\bibinfo{author}{\bibfnamefont{W.~F.} \bibnamefont{Brinkman}} \bibnamefont{and}
  \bibinfo{author}{\bibfnamefont{P.~W.} \bibnamefont{Anderson}},
  \bibinfo{journal}{Phys. Rev.} \textbf{\bibinfo{volume}{{A}8}},
  \bibinfo{pages}{2732} (\bibinfo{year}{1973}).

\bibitem[{\citenamefont{Pollanen et~al.}(2012)\citenamefont{Pollanen, Li,
  Collett, Gannon, Halperin, and Sauls}}]{pol12}
\bibinfo{author}{\bibfnamefont{J.}~\bibnamefont{Pollanen}},
  \bibinfo{author}{\bibfnamefont{J.~I.~A.} \bibnamefont{Li}},
  \bibinfo{author}{\bibfnamefont{C.~A.} \bibnamefont{Collett}},
  \bibinfo{author}{\bibfnamefont{W.~J.} \bibnamefont{Gannon}},
  \bibinfo{author}{\bibfnamefont{W.~P.} \bibnamefont{Halperin}},
  \bibnamefont{and} \bibinfo{author}{\bibfnamefont{J.~A.} \bibnamefont{Sauls}},
  \bibinfo{journal}{Nature} \textbf{\bibinfo{volume}{8}}, \bibinfo{pages}{317}
  (\bibinfo{year}{2012}).

\bibitem[{\citenamefont{Volovik}(2008)}]{vol08}
\bibinfo{author}{\bibfnamefont{G.~E.} \bibnamefont{Volovik}},
  \bibinfo{journal}{J. Low Temp. Phys.} \textbf{\bibinfo{volume}{150}},
  \bibinfo{pages}{453} (\bibinfo{year}{2008}).

\bibitem[{\citenamefont{Rainer and Vuorio}(1977)}]{rai77}
\bibinfo{author}{\bibfnamefont{D.}~\bibnamefont{Rainer}} \bibnamefont{and}
  \bibinfo{author}{\bibfnamefont{M.}~\bibnamefont{Vuorio}},
  \bibinfo{journal}{Journal Of Physics C-Solid State Physics}
  \textbf{\bibinfo{volume}{10}}, \bibinfo{pages}{3093} (\bibinfo{year}{1977}).

\bibitem[{\citenamefont{Surovtsev and Fomin}(2008)}]{sur08}
\bibinfo{author}{\bibfnamefont{E.~V.} \bibnamefont{Surovtsev}}
  \bibnamefont{and} \bibinfo{author}{\bibfnamefont{I.~A.} \bibnamefont{Fomin}},
  \bibinfo{journal}{Journal Of Low Temperature Physics}
  \textbf{\bibinfo{volume}{150}}, \bibinfo{pages}{487} (\bibinfo{year}{2008}).

\bibitem[{\citenamefont{Surovtsev}(2009)}]{sur09}
\bibinfo{author}{\bibfnamefont{E.~V.} \bibnamefont{Surovtsev}},
  \bibinfo{journal}{Journal Of Experimental And Theoretical Physics}
  \textbf{\bibinfo{volume}{108}}, \bibinfo{pages}{616} (\bibinfo{year}{2009}).

\bibitem[{\citenamefont{Kunimatsu et~al.}(2007)\citenamefont{Kunimatsu, Sato,
  Izumina, Matsubara, Sasaki, Kubota, Ishikawa, Mizusaki, and Bunkov}}]{kun07}
\bibinfo{author}{\bibfnamefont{T.}~\bibnamefont{Kunimatsu}},
  \bibinfo{author}{\bibfnamefont{T.}~\bibnamefont{Sato}},
  \bibinfo{author}{\bibfnamefont{K.}~\bibnamefont{Izumina}},
  \bibinfo{author}{\bibfnamefont{A.}~\bibnamefont{Matsubara}},
  \bibinfo{author}{\bibfnamefont{Y.}~\bibnamefont{Sasaki}},
  \bibinfo{author}{\bibfnamefont{M.}~\bibnamefont{Kubota}},
  \bibinfo{author}{\bibfnamefont{O.}~\bibnamefont{Ishikawa}},
  \bibinfo{author}{\bibfnamefont{T.}~\bibnamefont{Mizusaki}}, \bibnamefont{and}
  \bibinfo{author}{\bibfnamefont{Y.~M.} \bibnamefont{Bunkov}},
  \bibinfo{journal}{Jetp Letters} \textbf{\bibinfo{volume}{86}},
  \bibinfo{pages}{216} (\bibinfo{year}{2007}).

\bibitem[{\citenamefont{Meakin}(1998)}]{meakin98}
\bibinfo{author}{\bibfnamefont{P.}~\bibnamefont{Meakin}},
  \emph{\bibinfo{title}{{Fractals, scaling and growth far from equilibrium}}}
  (\bibinfo{publisher}{Cambridge University Press},
  \bibinfo{address}{Cambridge, UK}, \bibinfo{year}{1998}).

\bibitem[{\citenamefont{Porto and Parpia}(1999)}]{por99}
\bibinfo{author}{\bibfnamefont{J.~V.} \bibnamefont{Porto}} \bibnamefont{and}
  \bibinfo{author}{\bibfnamefont{J.~M.} \bibnamefont{Parpia}},
  \bibinfo{journal}{Phys. Rev. B} \textbf{\bibinfo{volume}{59}},
  \bibinfo{pages}{14583} (\bibinfo{year}{1999}).

\bibitem[{\citenamefont{Haard}(2001)}]{haa01}
\bibinfo{author}{\bibfnamefont{T.}~\bibnamefont{Haard}}, Ph.D. thesis,
  \bibinfo{school}{Northwestern University} (\bibinfo{year}{2001}).

\bibitem[{\citenamefont{Ma et~al.}(2000)\citenamefont{Ma, Roberts, Pr{\'e}vost,
  Jullien, and Scherer}}]{ma00}
\bibinfo{author}{\bibfnamefont{H.-S.} \bibnamefont{Ma}},
  \bibinfo{author}{\bibfnamefont{A.~P.} \bibnamefont{Roberts}},
  \bibinfo{author}{\bibfnamefont{J.-H.} \bibnamefont{Pr{\'e}vost}},
  \bibinfo{author}{\bibfnamefont{R.}~\bibnamefont{Jullien}}, \bibnamefont{and}
  \bibinfo{author}{\bibfnamefont{G.~W.} \bibnamefont{Scherer}},
  \bibinfo{journal}{Journal Of Non-Crystalline Solids}
  \textbf{\bibinfo{volume}{277}}, \bibinfo{pages}{127} (\bibinfo{year}{2000}).

\bibitem[{\citenamefont{Ma et~al.}(2002)\citenamefont{Ma, Jullien, and
  Scherer}}]{ma02b}
\bibinfo{author}{\bibfnamefont{H.-S.} \bibnamefont{Ma}},
  \bibinfo{author}{\bibfnamefont{R.}~\bibnamefont{Jullien}}, \bibnamefont{and}
  \bibinfo{author}{\bibfnamefont{G.}~\bibnamefont{Scherer}},
  \bibinfo{journal}{Phys Rev. E} \textbf{\bibinfo{volume}{65}},
  \bibinfo{pages}{041403} (\bibinfo{year}{2002}).

\bibitem[{\citenamefont{Larkin}(1970)}]{lar70}
\bibinfo{author}{\bibfnamefont{A.}~\bibnamefont{Larkin}},
  \bibinfo{journal}{Sov. Phys. JETP} \textbf{\bibinfo{volume}{31}},
  \bibinfo{pages}{784} (\bibinfo{year}{1970}).

\bibitem[{\citenamefont{Imry and Ma}(1975)}]{imr75}
\bibinfo{author}{\bibfnamefont{Y.}~\bibnamefont{Imry}} \bibnamefont{and}
  \bibinfo{author}{\bibfnamefont{S.}~\bibnamefont{Ma}}, \bibinfo{journal}{Phys.
  Rev. Lett.} \textbf{\bibinfo{volume}{35}}, \bibinfo{pages}{1399}
  (\bibinfo{year}{1975}).

\bibitem[{\citenamefont{Vollhardt and W\"olfle}(1990)}]{vollhardt90}
\bibinfo{author}{\bibfnamefont{D.}~\bibnamefont{Vollhardt}} \bibnamefont{and}
  \bibinfo{author}{\bibfnamefont{P.}~\bibnamefont{W\"olfle}},
  \emph{\bibinfo{title}{{The Superfluid Phases of $^3$He}}}
  (\bibinfo{publisher}{Taylor \& Francis}, \bibinfo{address}{New York},
  \bibinfo{year}{1990}).

\bibitem[{\citenamefont{Mermin and Stare}(1973)}]{mer73}
\bibinfo{author}{\bibfnamefont{N.~D.} \bibnamefont{Mermin}} \bibnamefont{and}
  \bibinfo{author}{\bibfnamefont{G.}~\bibnamefont{Stare}},
  \bibinfo{journal}{Phys. Rev. Lett.} \textbf{\bibinfo{volume}{30}},
  \bibinfo{pages}{1135} (\bibinfo{year}{1973}).

\bibitem[{\citenamefont{Brinkman and Smith}(1975)}]{bri75}
\bibinfo{author}{\bibfnamefont{W.~F.} \bibnamefont{Brinkman}} \bibnamefont{and}
  \bibinfo{author}{\bibfnamefont{H.}~\bibnamefont{Smith}},
  \bibinfo{journal}{Phys. Lett. A} \textbf{\bibinfo{volume}{51}},
  \bibinfo{pages}{449} (\bibinfo{year}{1975}).

\bibitem[{\citenamefont{Askhadullin
  et~al.}(2012{\natexlab{a}})\citenamefont{Askhadullin, Dmitriev, Krasnikhin,
  Martynov, Osipov, Senin, and Yudin}}]{ask12c}
\bibinfo{author}{\bibfnamefont{R.~S.} \bibnamefont{Askhadullin}},
  \bibinfo{author}{\bibfnamefont{V.~V.} \bibnamefont{Dmitriev}},
  \bibinfo{author}{\bibfnamefont{D.~A.} \bibnamefont{Krasnikhin}},
  \bibinfo{author}{\bibfnamefont{P.~N.} \bibnamefont{Martynov}},
  \bibinfo{author}{\bibfnamefont{A.~A.} \bibnamefont{Osipov}},
  \bibinfo{author}{\bibfnamefont{A.~A.} \bibnamefont{Senin}}, \bibnamefont{and}
  \bibinfo{author}{\bibfnamefont{A.~N.} \bibnamefont{Yudin}},
  \bibinfo{journal}{J. Phys. Conf. Ser.} \textbf{\bibinfo{volume}{to be
  published}}, \bibinfo{pages}{4 pages} (\bibinfo{year}{2012}{\natexlab{a}}).

\bibitem[{\citenamefont{Askhadullin
  et~al.}(2012{\natexlab{b}})\citenamefont{Askhadullin, Dmitriev, Krasnikhin,
  Martynov, Osipov, Senin, and Yudin}}]{ask12}
\bibinfo{author}{\bibfnamefont{R.~S.} \bibnamefont{Askhadullin}},
  \bibinfo{author}{\bibfnamefont{V.~V.} \bibnamefont{Dmitriev}},
  \bibinfo{author}{\bibfnamefont{D.~A.} \bibnamefont{Krasnikhin}},
  \bibinfo{author}{\bibfnamefont{P.~N.} \bibnamefont{Martynov}},
  \bibinfo{author}{\bibfnamefont{A.~A.} \bibnamefont{Osipov}},
  \bibinfo{author}{\bibfnamefont{A.~A.} \bibnamefont{Senin}}, \bibnamefont{and}
  \bibinfo{author}{\bibfnamefont{A.~N.} \bibnamefont{Yudin}},
  \bibinfo{journal}{Sov. Phys. JETP Lett.} \textbf{\bibinfo{volume}{95}},
  \bibinfo{pages}{326} (\bibinfo{year}{2012}{\natexlab{b}}).

\bibitem[{\citenamefont{Cross}(1975)}]{cro75}
\bibinfo{author}{\bibfnamefont{M.~C.} \bibnamefont{Cross}},
  \bibinfo{journal}{J. Low Temp. Phys.} \textbf{\bibinfo{volume}{21}},
  \bibinfo{pages}{525} (\bibinfo{year}{1975}).

\bibitem[{\citenamefont{Volovik and Mineev}(1981)}]{vol81}
\bibinfo{author}{\bibfnamefont{G.~E.} \bibnamefont{Volovik}} \bibnamefont{and}
  \bibinfo{author}{\bibfnamefont{V.~P.} \bibnamefont{Mineev}},
  \bibinfo{journal}{Zh. Eskp. Teor. Fiz.} \textbf{\bibinfo{volume}{81}},
  \bibinfo{pages}{867} (\bibinfo{year}{1981}), \bibinfo{note}{[ JETP , 60 , 276
  (1984)]}.

\bibitem[{\citenamefont{Choi and Muzikar}(1989)}]{cho89}
\bibinfo{author}{\bibfnamefont{C.}~\bibnamefont{Choi}} \bibnamefont{and}
  \bibinfo{author}{\bibfnamefont{P.}~\bibnamefont{Muzikar}},
  \bibinfo{journal}{Phys. Rev. Lett.} \textbf{\bibinfo{volume}{40}},
  \bibinfo{pages}{5144} (\bibinfo{year}{1989}).

\bibitem[{\citenamefont{Volovik}(1996)}]{vol96}
\bibinfo{author}{\bibfnamefont{G.~E.} \bibnamefont{Volovik}},
  \bibinfo{journal}{Sov. Phys. JETP Lett.} \textbf{\bibinfo{volume}{63}},
  \bibinfo{pages}{281} (\bibinfo{year}{1996}).

\bibitem[{\citenamefont{Pollanen et~al.}(2007)\citenamefont{Pollanen,
  Blinstein, Choi, Davis, Lippman, Lurio, and Halperin}}]{pol07}
\bibinfo{author}{\bibfnamefont{J.}~\bibnamefont{Pollanen}},
  \bibinfo{author}{\bibfnamefont{S.}~\bibnamefont{Blinstein}},
  \bibinfo{author}{\bibfnamefont{H.}~\bibnamefont{Choi}},
  \bibinfo{author}{\bibfnamefont{J.~P.} \bibnamefont{Davis}},
  \bibinfo{author}{\bibfnamefont{T.~M.} \bibnamefont{Lippman}},
  \bibinfo{author}{\bibfnamefont{L.}~\bibnamefont{Lurio}}, \bibnamefont{and}
  \bibinfo{author}{\bibfnamefont{W.}~\bibnamefont{Halperin}},
  \bibinfo{journal}{J. Low Temp. Phys.} \textbf{\bibinfo{volume}{148}},
  \bibinfo{pages}{579} (\bibinfo{year}{2007}).

\bibitem[{\citenamefont{Sauls and Sharma}(2010)}]{sau10}
\bibinfo{author}{\bibfnamefont{J.~A.} \bibnamefont{Sauls}} \bibnamefont{and}
  \bibinfo{author}{\bibfnamefont{P.}~\bibnamefont{Sharma}},
  \bibinfo{journal}{New J. Phys.} \textbf{\bibinfo{volume}{12}},
  \bibinfo{pages}{083056} (\bibinfo{year}{2010}).

\bibitem[{\citenamefont{Abrikosov and Gorkov}(1961)}]{abr61}
\bibinfo{author}{\bibfnamefont{A.~A.} \bibnamefont{Abrikosov}}
  \bibnamefont{and} \bibinfo{author}{\bibfnamefont{L.~P.}
  \bibnamefont{Gorkov}}, \bibinfo{journal}{Sov. Phys. JETP}
  \textbf{\bibinfo{volume}{12}}, \bibinfo{pages}{1243} (\bibinfo{year}{1961}).

\bibitem[{\citenamefont{Sauls and Sharma}(2003)}]{sau03}
\bibinfo{author}{\bibfnamefont{J.~A.} \bibnamefont{Sauls}} \bibnamefont{and}
  \bibinfo{author}{\bibfnamefont{P.}~\bibnamefont{Sharma}},
  \bibinfo{journal}{Phys. Rev. B} \textbf{\bibinfo{volume}{68}},
  \bibinfo{pages}{224502} (\bibinfo{year}{2003}).

\bibitem[{\citenamefont{Thuneberg}(1987)}]{thu87}
\bibinfo{author}{\bibfnamefont{E.}~\bibnamefont{Thuneberg}},
  \bibinfo{journal}{Phys. Rev. B} \textbf{\bibinfo{volume}{36}},
  \bibinfo{pages}{3583} (\bibinfo{year}{1987}).

\bibitem[{\citenamefont{Li}(2010)}]{li13}
\bibinfo{author}{\bibfnamefont{J.~I.~A.} \bibnamefont{Li}} \bibnamefont{and}
  \bibinfo{author}{\bibfnamefont{A.M.}~\bibnamefont{Zimmerman}} \bibnamefont{and}
  \bibinfo{author}{\bibfnamefont{J.}~\bibnamefont{Pollanen}} \bibnamefont{and}
  \bibinfo{author}{\bibfnamefont{C.~A.}~\bibnamefont{Collett}}\bibnamefont{and}
  \bibinfo{author}{\bibfnamefont{W.~J.}~\bibnamefont{Gannon}} \bibnamefont{and}
  \bibinfo{author}{\bibfnamefont{W.~P.}~\bibnamefont{Halperin}},
  \bibinfo{journal}{J. Low Temp. Phys.} %\textbf{\bibinfo{volume}{X}},
  \bibinfo{pages}{submitted}[\bibinfo{journal}{arXiv}
  \textbf{\bibinfo{volume}{1306.5783}}, \bibinfo{pages}{6}]
 (\bibinfo{year}{2013}).

\end{thebibliography}
%

%
\end{document}